\documentclass[%
 reprint,
 amsmath,amssymb,
 aps,
floatfix,
]{revtex4-2}

\usepackage{graphicx}
\usepackage{dcolumn}
\usepackage{bm}
\usepackage{subcaption}
\usepackage{MnSymbol}  
\usepackage[dvipsnames]{xcolor}
\usepackage{changepage}
\def\simin{\ensuremath{\;\tilde{\in}\;}}  

\begin{document}

\preprint{APS/123-QED}

\title{Illuminating Systematic Trends in Nuclear Data with Generative Machine Learning Models}

\author{Jordan M. R. Fox}
\email{jfox@anl.gov}
\affiliation{Argonne National Laboratory, Lemont, Illinois, USA\\ San Diego State University, San Diego, California, USA}

\author{Kyle A. Wendt}
\email{wendt6@llnl.gov}
\affiliation{Lawrence Livermore National Laboratory, Livermore, California, USA}%

\date{\today}

\begin{abstract}
We introduce a novel method for studying systematic trends in nuclear reaction data using generative adversarial networks. 
Libraries of nuclear cross section evaluations exhibit intricate systematic trends across the nuclear landscape, and predictive models capable of reproducing and analyzing these trends are valuable for many applications.
We have developed a predictive model using deep generative adversarial networks to learn trends from the inelastic neutron scattering channel of TENDL for even-even nuclei. 
The system predicts cross sections based on adding/subtracting particles to/from the target nucleus. 
It can thus help identify cross sections that break from expected trends and predict beyond the limit of current experiments. 
Our model can produce good predictions for cross section curves for many nuclides, and it is most robust near the line of stability. 
Furthermore, we can generate ensembles of predictions to leverage different correlations and estimate model uncertainty. 
This research marks an important first step in the development of generative AI models to nuclear cross-section libraries.
\end{abstract}

\maketitle

\section{\label{sec:intro}Introduction}

\subsection{\label{sec:ncse}Nuclear cross section evaluations}
\newcommand{\fakecite}[1]{\textbf{\textbackslash{}cite}\{#1\}}
Simple patterns and trends have long been a staple in the phenomenology of atomic nuclei. 
The values that nuclear masses take as functions of the number of protons (Z) and neutrons (N) are one of the best-studied cases of such trends. 
Despite the rich and complex quantum many-body structure that enters a precise description of nuclear ground states, their masses are well described by simple semi-empirical formulae, such as a liquid drop model \cite{Weizsacker:1935bkz}, which captures the dependence of nuclear masses in terms of total number of nucleons ($N + Z$), asymmetry between protons and neutrons ($N - Z$), and corrections for effects such as pairing. 
More recent studies exploit trends related to static properties (nuclear skins, electromagnetic transition strengths, etc.) to build ensembles of nuclear models (or ensembles of parameterizations of a specific model) and connect predictions to experimental measurements \cite{Hagen:2015yea, PhysRevLett.120.172702}.  
This work has a slightly different aim. Instead of looking at relatively simple data with prominent trends or looking at the pattern between purely theoretical models, we seek to learn about trends hidden in libraries of evaluated nuclear reaction cross sections.

Evaluated nuclear reaction data, such as the ENDF\cite{BROWN20181}, JEFF\cite{JEFF}, or TENDL\cite{KONING20191} libraries (non-exhaustively), represents the interface between nuclear physics and other sciences and engineering that depend on nuclear physics.  The consumers of nuclear data span from astrophysical simulations and high energy physics (through detector/background design and modeling) to nuclear power, safety, and radiological medicine \cite{FENG202112}.  These evaluated libraries are a fusion of experimental data and theoretical models that aim to give a comprehensive picture of scattering processes for as much of the nuclear chart as possible.  These libraries contain complicated systematic trends due to fundamental nuclear physics, but they can be difficult to study due to their tremendous size, density of information, and the complexity of the evaluations.  From a theoretical perspective, the whole libraries are not presently well described by a singular model, and instead, a patchwork of different models are combined locally (in proton and neutron numbers) across the chart, obfuscating these trends.  Our goal for this work is to develop a machine learning system to facilitate analysis of trends in cross sections and, using existing evaluation libraries, predict cross sections beyond experimental barriers.

Presently, we focus on the TENDL library \cite{KONING20191} and the inelastic neutron scattering channel $(n,n')$ in particular, with resonances excluded; this choice was made because of the relative simplicity of correlations in this channel. 
Because of odd-even staggering, the well-known behavior of systematic trends in nuclei to oscillate with parity of proton/neutron numbers $Z, N$, we only work with even-even nuclei in the present study. 
Overall, these simplifications make the learning task easier, allowing for more interpretable results, but it still presents a significant challenge.
This work is a foundation for developing more sophisticated models.  

This data set has many different types of correlations (i.e., at different scales over the chart of nuclides), making it a ripe target for data science and machine learning.
For instance, a predictive model must learn correlations in the cross sections between values of scattering energy, correlations between nuclei with similar numbers of constituent particles, and (in future development) correlations between cross sections in different reaction channels. 
In many cases, these cross section curves exhibit interesting structure at low energy (0-10 MeV) and a relatively predictable tail at higher energies (10-20 MeV). 
One of the most obvious, yet ultimately challenging, qualifications of a predictive model is correctly reproducing the low energy structure.

Recent advances in artificial intelligence applications to nuclear physics have been numerous and broad in scope \cite{bedaque2021ai, Boehnlein_colloquium_2022, He_Ma_Pang_Song_Zhou_2023, He_Li_Ma_Niu_Pei_Zhang_2023,PhysRevD.100.034515}.  
The present work is most similar to data-driven methods for nuclear masses \cite{Gao_Wang_Lu_Li_Shen_Liu_2021, PhysRevC.106.L021301, Niu_Liang_2022} and error detection in engineering libraries \cite{Grechanuk_Rising_Palmer_2021},
but significant progress has also been made on ML (including neural network) applications to the many-fermion problem \cite{Adams_Carleo_Lovato_Rocco_2021, Rigo_Hall_Hjorth-Jensen_Lovato_Pederiva_2023,Lovato_Adams_Carleo_Rocco_2022}. 

\subsection{\label{sec:overview}Problem statement \& approach}

\begin{figure}
    \centering
    \includegraphics[scale=0.075]{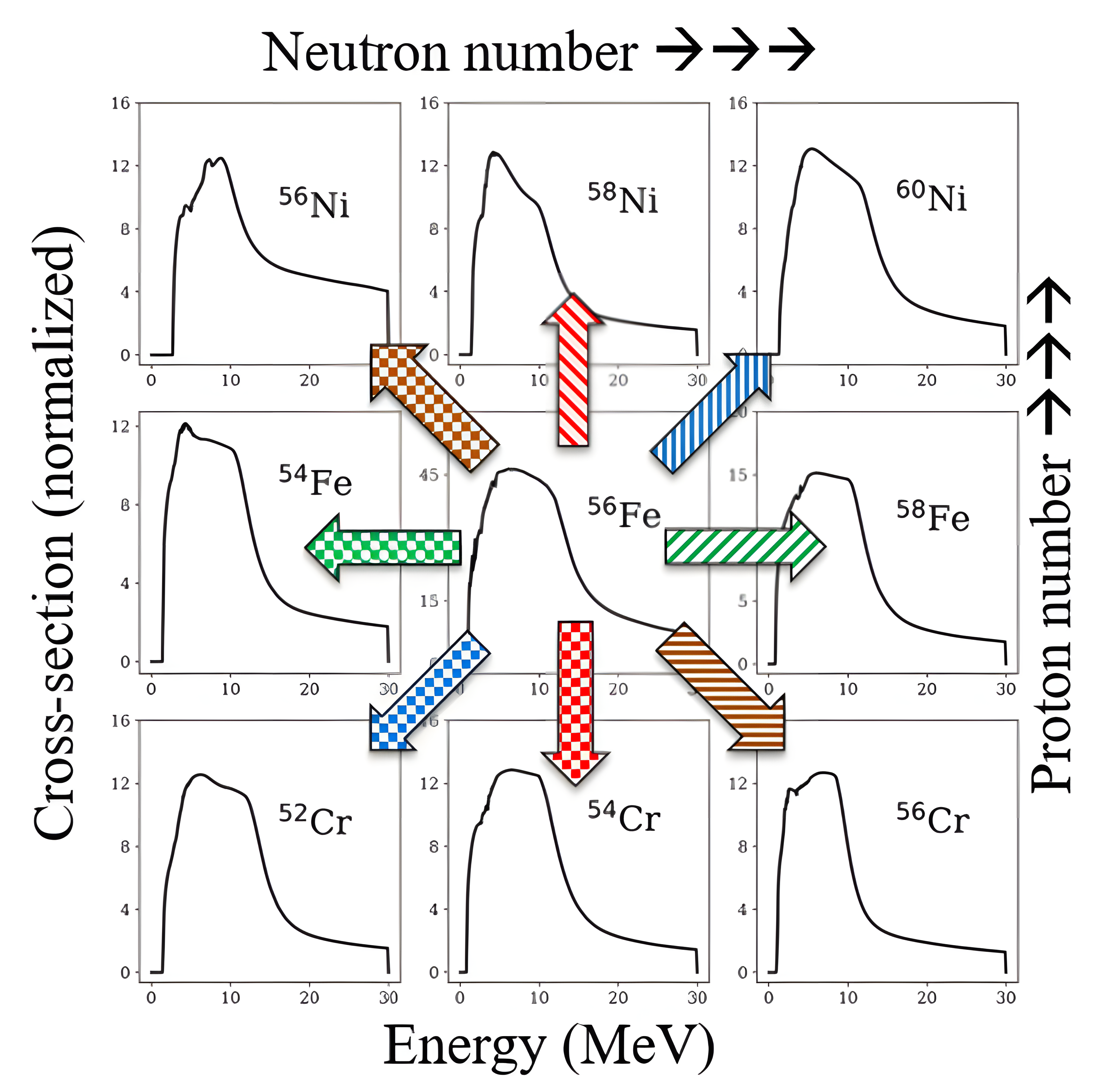}
    \caption{Our model predicts the cross section curve for one nuclide given that of a direct neighbor. On the chart, each nuclide has eight direct neighbors. Each arrow in the figure corresponds to a part of the larger system tasked with learning trends in that direction. }
    \label{fig:tendl_grid}
\end{figure}

Here we describe the problem of interest and the general approach. 
Consider a small section of the chart of nuclides, like that shown in Fig.~\ref{fig:tendl_grid}, with only even-even nuclides. 
Each tile shows the $(n,n')$ cross section curve (in this case, with an arbitrary normalization), and each of the eight arrows corresponds to a separate \textbf{translation} task learned by our model.
For example, beginning with the curve for $^{56}$Fe, one neural network will map to the cross section curve for $^{58}$Fe. 
We can identify this translation generally as a direction on the chart: $(N,Z)\rightarrow(N+2,Z)$. 
Likewise, another neural network will map from $^{56}$Fe to $^{60}$Ni; that is, $(N,Z) \rightarrow(N+2,Z+2)$. 
The goal is for each of the eight networks to learn a general mapping on the chart and predict any curve from one of its neighbors. 
The collection of networks ultimately contains information about how the cross section curve changes globally with respect to $N$ and $Z$ and may be used to gain further insight.  

The full predictive system, made up of this set of networks, is thus an ensemble.
One cross section can be predicted by starting at any other nuclide and convolving the necessary translations.
We can make predictions for cross sections of nuclides inside and outside the training region by combining predictions using many paths across the chart, allowing for all parts of the model to be utilized in each prediction.

The next section discusses the data encoding process (subsec.~\ref{sec:encoding}), the adversarial model responsible for translating the cross sections (subsec.~\ref{sec:gans}).
Section \ref{sec:predictions} explores several ways the larger model can be evaluated and used to predict cross sections.

\section{A Predictive AI model for cross section curves}

\subsection{\label{sec:encoding}Data encoding with a convolutional variational auto-encoder (VAE)} 

The first step in our procedure, typical for machine learning tasks, is to find an ideal representation of the data. 
This includes normalization and an \textit{encoding} to simplify correlations in the data and reduce dimension. 
We first standardize all cross section curves to lie on an energy domain of $0-30$ MeV in 256 bins and normalize them individually to a maximum of 1. 
This way, we factor out the task of predicting amplitudes and just focus on the shape of the curve. 

Our chosen method for encoding the cross section curves is a type of neural network model called a variational auto-encoder (VAE) \cite{kingma2013_vae}. 
An \textit{auto-encoder} (AE) \cite{rumelhart1986learning} is a powerful method of data encoding that is tantamount to a nonlinear principal component analysis \cite{kramer1991nonlinear}. 
When doing standard linear principal component analysis via singular value decomposition (SVD), one finds orthogonal principal components of data, and the whole data set can be represented using those components as a basis. Equation \ref{eq:svd} shows this approach; $\Sigma $ is a matrix with the standardized cross sections, $\sigma_{(N,Z)}$,  as columns, for all included nuclides $(N,Z)$. 

\begin{equation}
    \Sigma = U D W^T
    \label{eq:svd}
\end{equation}

The results of SVD provide the set of principal components $W$ present in the data, their singular values $D$ (a diagonal matrix), and the necessary linear combinations $U$. One may then perform dimensionality reduction by ignoring those components with the smallest corresponding singular values, resulting in efficient encoding at the expense of some errors.  

\begin{figure}
    \centering
    \includegraphics[scale=0.3]{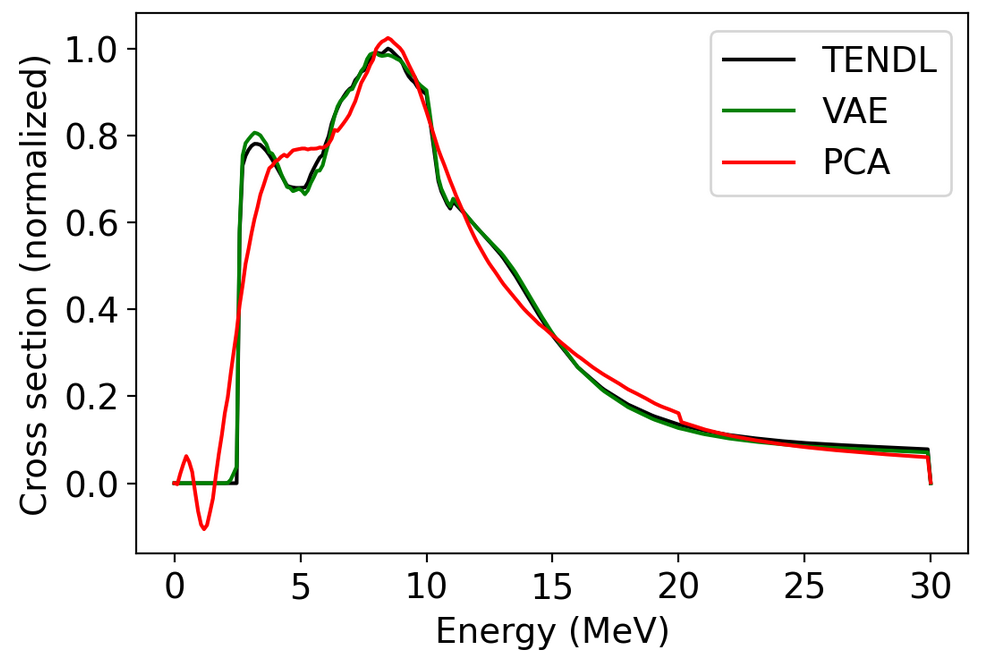}
    \caption{The variational auto-encoder outperforms a linear principle component analysis with the same number of variables. }
    \label{fig:linear_pca_result}
\end{figure}

However, linear PCA includes some crucial assumptions: the data is a linear combination of explanatory variables, all explanatory variables have been observed roughly the same amount, correlations follow a Gaussian distribution, the data is primarily unimodal, etc. 
Cross section evaluation libraries are the result of different models, experimental techniques, and resolutions; to an extent, none of these assumptions hold. 
Furthermore, cross sections are strictly positive, and using linear PCA to reduce dimension inevitably produces negative cross sections; these would then have to be filtered of negative values, which is an irreversible process.
For these reasons, we did not use linear PCA for encoding; Fig.~\ref{fig:linear_pca_result} illustrates the improved compression capability of the VAE over linear PCA (using the same number of variables). 

The VAE is a deep neural network mapping from data set $\Sigma$ to itself, and the smallest layer inside the network is the \textbf{latent space} representation of the input. 
The model is split into an \textbf{encoder} part $E$ before the latent layer and a \textbf{decoder} part $E^+$ after the latent layer (borrowing notation from the Moore-Penrose pseudoinverse). 
The VAE model can be nicely represented as a function convolution,  $E^+(E(\sigma_{(N,Z)})) = \sigma'_{(N,Z)}$, where $\sigma'_{(N,Z)}$ is the reconstructed cross section.
Denoting weights $\theta$ and $\theta^+$ for $E$ and $E^+$, respectively, the standard loss function $\mathcal{L}_\text{AE}$ penalizes reconstruction error as shown in Eq.~\ref{eq:aeloss}.
The decoder, $E^+$, approaches $ E^{-1}$ as the loss function is minimized; the encoder and decoder transform to and from the latent space. 
Implicit in Eq.~\ref{eq:aeloss} is choice of a batch of $N_b$ data from the total data set $\Sigma$, indexed by $i$. This is a technical detail; for now one may take $N_b=|\Sigma|$ and let $i$ replace $(N,Z)$ for brevity.

\begin{equation}\label{eq:aeloss}
    \mathcal{L}_\text{AE}(\theta,\theta^+;\Sigma) = \frac{1}{N_b} \sum_{i=1}^{N_b} |E^+_{\theta^+} (E_{\theta}(\sigma_i)) - \sigma_i |^2
\end{equation}

Since we will use the encoded data as input for another model, having a smooth and efficient encoding is a high priority; the use of variational Bayes (VB) \cite{tran2021practical} in the VAE gives us just that. 
VB generally refers to statistical methods that approximate a probability distribution with a conditional (or joint) one, often using Kullback-Leibler divergence (KLD) as a distance metric.
In essence, the VAE is a mapping of the VB problem onto a standard auto-encoder structure.
In what is known as the \textbf{reparameterization trick} \cite{kingma2013_vae}, the encoder maps to a mean and variance of a multivariate normal distribution, rather than points. 
In other words, the latent representation of $\sigma_i$ used to train the VAE is $E(\sigma_i) = (m_{\zeta,i} , s^2_{\zeta,i})$, where $m_{\zeta,i}$ and $s^2_{\zeta,i}$ correspond to a normal mean and variance. 
At each training step, one generates a new random sample $\zeta' = \mathcal{N}(m_{\zeta,i},s_{\zeta,i}^2)$ which is decoded to satisfy $E^+(\zeta_i')=\sigma_i'$.
Kullback-Leibler divergence is introduced to the loss function to bring $\mathcal{N}(m_{\zeta,i},s_{\zeta,i}^2)$ toward $\mathcal{N}(0,1)$, a unit-normal distribution.
The total VAE loss function is thus  
\begin{equation}
\begin{aligned}
    \mathcal{L}_\text{VAE}(\theta,\theta^+;\Sigma, \beta) &=  - \frac{\beta}{N_b} \sum_{i=1}^{N_b} \text{KLD} \left[ \mathcal{N}(m_{\zeta,i} , s^2_{\zeta,i}) || \mathcal{N}(0 , 1) \right]\\ &+ \mathcal{L}_\text{AE}(\theta,\theta^+;\Sigma),    
\end{aligned}
\end{equation}
where 
\begin{equation}
    \text{KLD} \left[ \mathcal{N}(m_{\zeta,i} , s^2_{\zeta,i}) || \mathcal{N}(0 , 1) \right] = \frac{1}{2} \left[  1 + \log s_{\zeta,i}^2 - m_{\zeta,i}^2 - s_{\zeta,i}^2 \right] .
\end{equation}
After training, the random sampling is removed, so the encoding is deterministic $E(\sigma_i)=m_{\zeta,i}$; we then denote the encoded data $m_{\zeta,i} = \zeta_i = E(\sigma_i)$ for simplicity. 

The scale of the KLD loss term relative to $\mathcal{L}_\text{AE}$ is controlled by the hyperparameter $\beta \in [0,1]$, which, in practice, may be tweaked during training to get better convergence. With a sufficiently complex neural network, it is easy to find (via numerical optimization with stochastic gradient descent) a minimization of KLD with large reconstruction error: in other words, the latent space representations are nearly unit normal, but $E$ and $E^+$ are not good approximations to the distributions they should be learning. To help this, a good trick is to schedule the KLD $\beta$  to oscillate from 0 to 1 several times before ultimately fixing it to 1 \cite{higgins2017betavae}. 
In our case, this resulted in good convergence and is especially easy to implement in code. 

In addition, our VAE's encoding and decoding layers are fully convolutional.  
Cross sections have strong short-range correlations and weak long-range correlations; in other words, the covariance matrix is diagonally dominant. 
We leverage this property by using convolutional layers in $E$ and $E^+$, learning multi-scale correlations; each consecutive hidden layer is responsible for a slightly larger or smaller correlation length.
This is similar to AI models used for image recognition: long-range correlations are not ignored, but the relative importance of short-range correlations is built directly into the network architecture. 
A convolutional neural network is likely superior to the usual densely connected feed-forward neural network for data dominated by local correlations, especially for a use-case like encoding.  
This is especially true in complicated problems where larger neural networks are employed; for a fixed number of layers, the convolutional network has fewer parameters than the dense one and, thus, is generally easier to train.
See \cite{goodfellow2016deep, o2015introduction} for more information on deep convolutional networks.

The VAE was trained using standard stochastic gradient descent and converged with sufficient accuracy. Once trained, the encoder and decoder networks were frozen and saved to be used in tandem with the predictive models discussed in the next section.  

\subsection{\label{sec:gans}Data translation with cycle-consistent generative adversarial networks (cycleGANs)}

\subsubsection{Model construction}

Research in deep generative learning, and the subset of it which is adversarial, has expanded and matured significantly in the years since the first GAN work was published \cite{gan_goodfellow}. 
We have developed our architecture based on experience with the relevant physics, but there are many different types of GAN models one may consider for this task \cite{stylegan_nvidia,dcgan_radford,cdd_gonzalezgarcia,multimodal_huang,multimodal_zhu}. 
Some background on simpler GAN constructions is provided in Appendix \ref{app:simplegans}.

We employ a modified version of the \textbf{cycleGAN} first introduced by Zhu et al.\cite{cyclegan_zhu}.
Before explaining the cycleGAN, we introduce some standards and notation for clarity.
The chart of nuclides is oriented with $N$ on the horizontal axis and $Z$ on the vertical; the reader is encouraged to picture the cycleGAN taking steps on the chart as illustrated in Fig.~\ref{fig:tendl_grid}.
First, for individual directions on the chart of nuclides (cardinals and diagonals), we use the symbol \mbox{$d \in \{ \uparrow, \nearrow, \rightarrow, \searrow, \downarrow, \swarrow, \leftarrow, \nwarrow \}$}. For example, ``${(N,Z) + \searrow}$'' is shorthand for $(N,Z)+(2,-2)=(N+2,Z-2)$.
Let $\bar{d}$ denote the opposite direction of $d$.
Second, we use the symbol $r$ for ``rule'', meaning each pair of co-inverse directions, so \mbox{$r \in \{ \updownarrow, \neswarrow, \leftrightarrow, \nwsearrow \}$}.
The $\leftrightarrow$ rule corresponds to adding/subtracting neutron pairs, the $\updownarrow$ rule to adding/subtracting proton pairs, the $\neswarrow$ rule to adding/subtracting both proton and neutron pairs, and the $\nwsearrow$ rule to swapping a proton pair for neutron pair and vice versa. 
\begin{equation}
    \begin{aligned}
        M_\leftrightarrow &= \{T_\rightarrow,T_\leftarrow,D_\leftrightarrow \} \\
        M_\updownarrow &= \{T_\uparrow,T_\downarrow,D_\updownarrow \} \\
        M_{\neswarrow} &= \{T_\nearrow,T_\swarrow,D_\neswarrow \} \\
        M_{\nwsearrow} &= \{T_\searrow,T_\nwarrow,D_\nwsearrow \} \\
    \end{aligned}
    \label{eq:model_sets_simplified}
\end{equation}
The entire predictive model is made up of four cycleGANs, not including the VAE; each one, denoted $M_r$, has a corresponding rule as shown in Eq.~\ref{eq:model_sets_simplified}. One cycleGAN consists of three networks: two translators $T_d$ and $T_{\bar{d}}$, responsible for translation along the rule, and one discriminator $D_r$, responsible for validating pairs of neighboring (encoded) cross sections along the rule.
These networks are all densely-connected feed-forward neural networks. 

\begin{figure*}
\begin{subfigure}{\textwidth}
\includegraphics[scale=0.2]{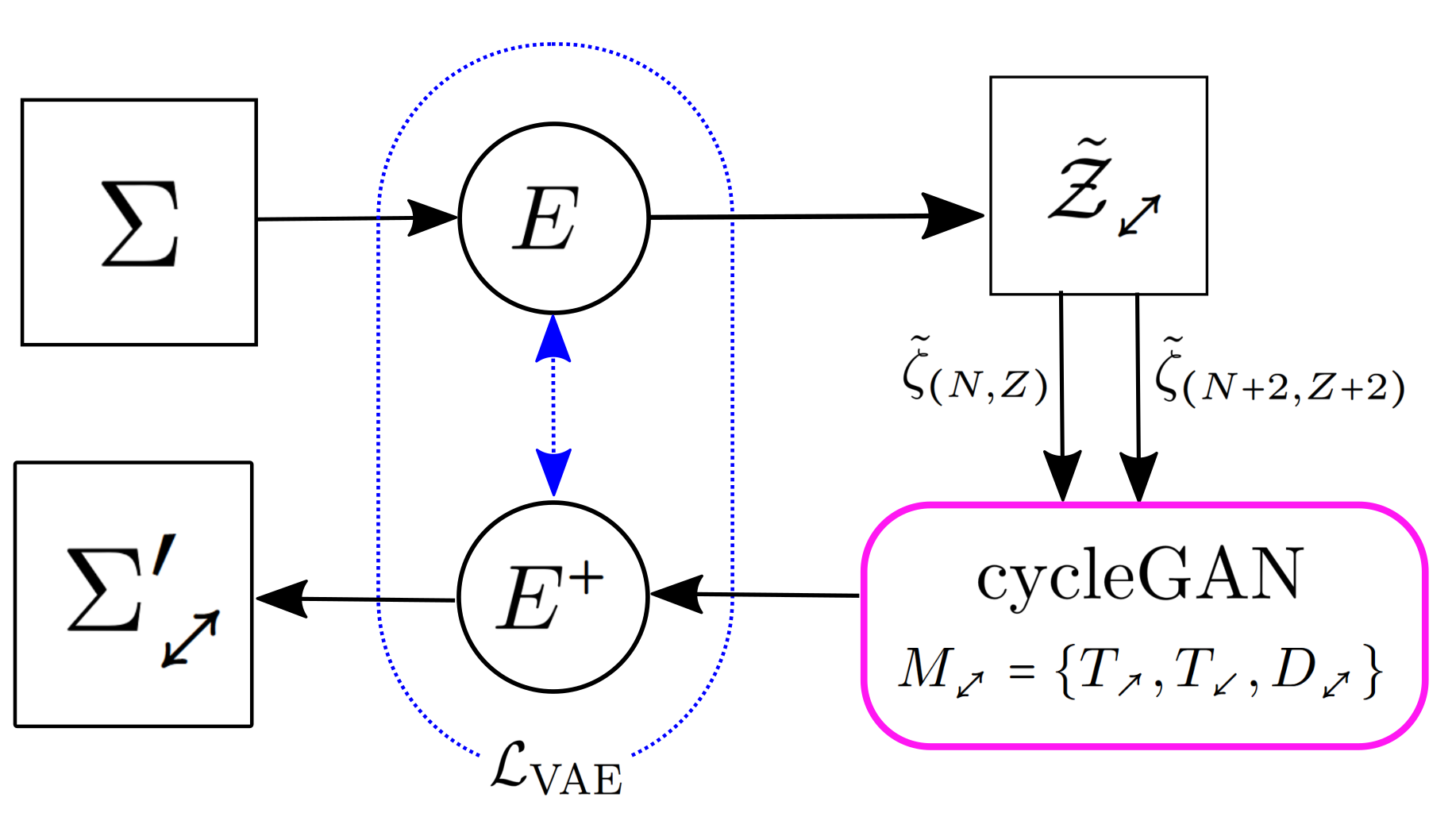}
\caption{\label{fig:flow1}
The normalized cross section data $\Sigma$ is first encoded $E$ and appended with numbers $(N,Z)$ to form the training set $\tilde{\mathcal{Z}}_\neswarrow$, which is fed to the cycleGAN. The corresponding encoded predictions $\tilde{\mathcal{Z}}'_\neswarrow$ are then decoded ($E^+$), which provides predictions of cross section curves $\Sigma'_\neswarrow$.}
\end{subfigure}
\begin{subfigure}{\textwidth}
\includegraphics[scale=0.2]{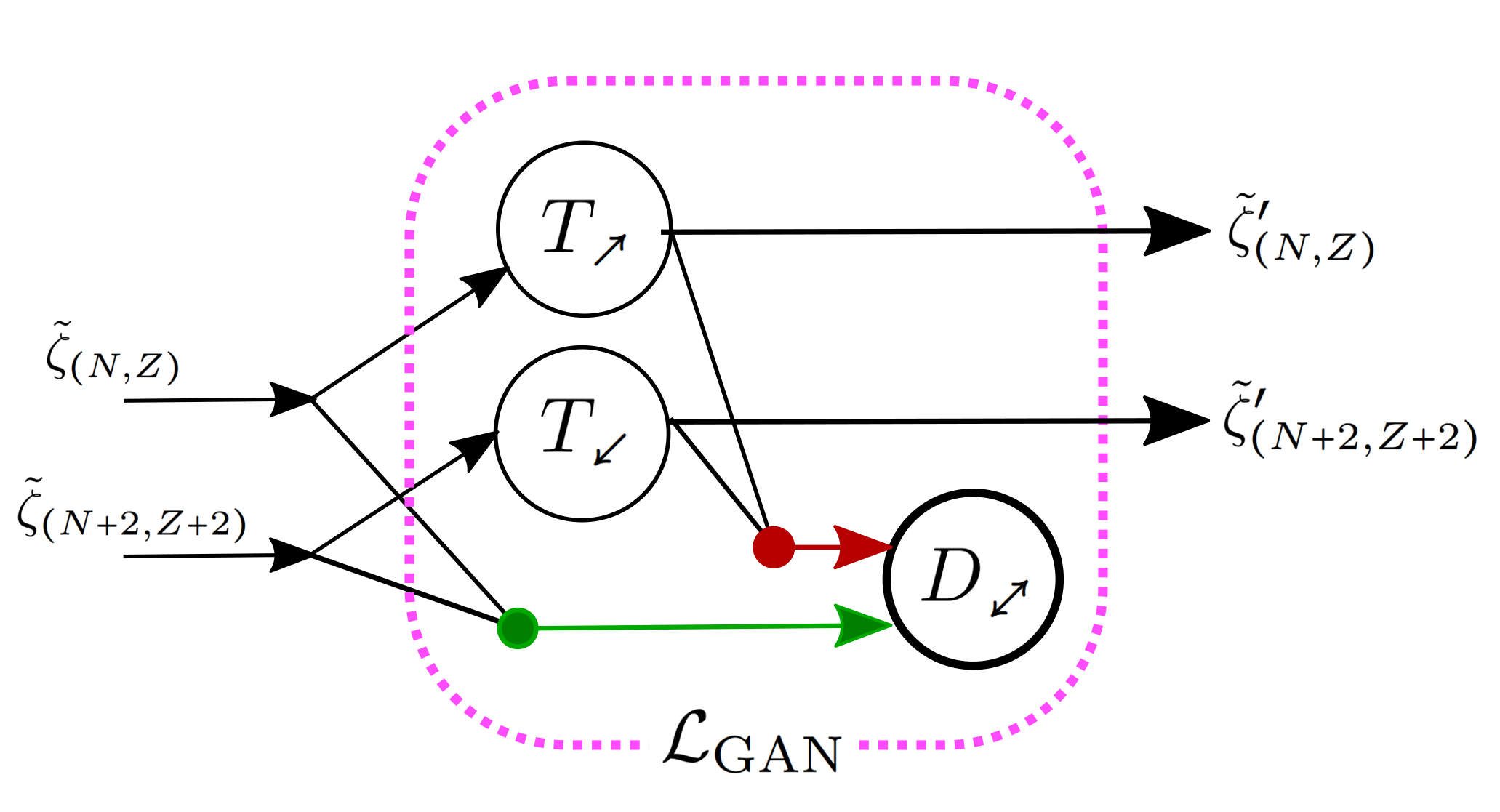}
\caption{\label{fig:flow2}
The cycleGAN model accepts the pair of encoded vectors, $\tilde{\zeta}_{(N,Z)}$ and $\tilde{\zeta}_{(N+2,Z+2)}$ without loss of generality. The pair is concatenated to provide the discriminator $D_\neswarrow$ a sample of ``real'' data. Each vector is then translated by the respective network $T_\nearrow, T_\swarrow$; this pair is concatenated to provide the discriminator with a sample of ``fake'' data. The GAN outputs the predicted vectors $\tilde{\zeta}'_{(N,Z)}$ and $\tilde{\zeta}'_{(N+2,Z+2)}$.}
\end{subfigure}
\caption{Flowcharts illustrating prediction procedure using the VAE and cycleGAN. The notation shown corresponds to the $\neswarrow$ rule on the nuclear chart; without loss of generality, each model has the same structure ($\updownarrow, \neswarrow, \leftrightarrow, \nwsearrow$).}
\label{fig:flowcharts}
\end{figure*}

As an example, consider the translation of cross sections between nuclide $(N,Z)$ and its neighbor ${(N+2,Z+2)}$. 
One step corresponds to the direction $\nearrow$ on our chart of nuclides. To translate the cross section, the curve $\sigma_{(N,Z)}$ is encoded, $E(\sigma_{(N,Z)})=\zeta_{(N,Z)}$, then translated, $T_\nearrow(\zeta_{(N,Z)}) = \zeta'_{(N+2,Z+2)}$, and finally decoded to produce the new curve, $E^+(\zeta'_{(N+2,Z+2)})=\sigma'_{(N+2,Z+2)}$.
This constitutes the prediction of our model for the cross section of nuclide ${(N+2,Z+2)}$ given that of ${(N,Z)}$.
Likewise, the same cycleGAN $M_\neswarrow$ can do the complimentary prediction: the curve $\sigma_{(N+2,Z+2)}$ is encoded, $E(\sigma_{(N+2,Z+2)})=\zeta_{(N+2,Z+2)}$, then translated, $T_\swarrow(\zeta_{(N+2,Z+2)}) = \zeta'_{(N,Z)}$, and finally decoded, $E^+(\zeta'_{(N,Z)})=\sigma'_{(N,Z)}$.
The discriminator $D_\neswarrow$ judges a pair of encoded cross sections, and is trained such that $D_\neswarrow(\zeta_{(N,Z)}, \zeta_{(N+2,Z+2)})=1$ (pass) and $D_\neswarrow(\zeta'_{(N,Z)}, \zeta'_{(N+2,Z+2)})=0$ (fail). 
That is, the discriminator gives the probability that the given pair of vectors match identifiable correlations in the rule.
This process is illustrated in Fig.~\ref{fig:flow2}.
Furthermore, we may invoke cycle-consistency: $T_\swarrow(\zeta'_{(N+2,Z+2)})) = \zeta''_{(N,Z)}$ brings us back to the starting nuclide. 
The differences between $\zeta_{(N,Z)}$, $\zeta'_{(N,Z)}$, and $\zeta''_{(N,Z)}$ ultimately depend on training and the loss function, which is discussed in the next section.

Each translator maps between encoded cross section vectors, albeit with one important modification. 
Each encoded vector $\zeta_{(N, Z)}$ is appended with normalized values of proton and neutron number $N, Z$; we denote this augmented form as $\tilde{\zeta}_{(N, Z)}$.
As such, the dimension of the translator input and output is equal to the VAE latent dimension plus 2.
This augmentation is done to provide additional support for the network to learn changes in cross sections, particularly in cases where $\sigma_{(N,Z)}$ closely resembles $\sigma_{(N,Z)\pm d}$. 
While we did not find it necessary to introduce more variables, other neural network models for nuclear data \cite{PhysRevC.106.L021301} have shown clear improvements in convergence when redundant physical information is included as additional features ($N-Z$,$A^{1/3}$, etc.). 

Figure \ref{fig:flowcharts} illustrates the workflow described in this section. In Fig.~\ref{fig:flow1}, cross section data from $\Sigma$ is encoded and augmented to produce the training data set $\tilde{\mathcal{Z}}_r $ which contains pairs $  {(\tilde{\zeta}_{(N,Z)} , \tilde{\zeta}_{(N,Z)+d})} \}$. This encoded cross section data is fed to the cycleGAN, which in turn predicts a set of cross section curves given the trends on that rule, $\Sigma_r$. In Fig.~\ref{fig:flow2}, we see the process described above.

\subsubsection{Model optimization}

Each cycleGAN $M_r=\{  T_d , T_{\bar{d}}, D_r\}$ is trained independently on a training set constructed according to the corresponding rule, $r=\updownarrow,\neswarrow,\leftrightarrow,\nwsearrow$.
Preparation of training data for each cycleGAN takes two steps. 
First, we use the VAE to encode each element of $\Sigma$, $E(\sigma_{(N, Z)}) = \zeta_{(N, Z)}$, which forms the full encoded data set $\mathcal{Z}$. 
Second, as mentioned in the previous section, each vector $\zeta_{(N, Z)}$ is appended with normalized values of proton and neutron number and denoted $\tilde{\zeta}_{(N, Z)}$.
Finally, every pair of neighboring nuclides along a given rule are paired together, as shown in Eq.~\ref{eq:trainingdata}.
We denote the resulting form of our training data as $\tilde{\mathcal{Z}}_r$.

\begin{equation}\label{eq:trainingdata}
    \begin{aligned}
       \tilde{\mathcal{Z}}_\leftrightarrow &= \{ (\tilde{\zeta}_{(N,Z)} , \tilde{\zeta}_{(N+2,Z)} )\} \\
       \tilde{\mathcal{Z}}_\updownarrow &= \{ (\tilde{\zeta}_{(N,Z)} , \tilde{\zeta}_{(N,Z+2)} )\} \\
       \tilde{\mathcal{Z}}_\neswarrow &= \{ (\tilde{\zeta}_{(N,Z)} , \tilde{\zeta}_{(N+2,Z+2)} )\} \\
       \tilde{\mathcal{Z}}_\nwsearrow &= \{ (\tilde{\zeta}_{(N,Z)} , \tilde{\zeta}_{(N+2,Z-2)} )\} \\
    \end{aligned}
\end{equation}

We employ four loss components: prediction loss, cycle loss, discriminator loss, and adversarial loss.
\begin{itemize}
    \item Prediction loss, Eq.~\ref{eqn:prediction_loss}, is equal to the traditional mean-absolute-error between the translator predictions and target cross section curve. 
Without this constraint, the total loss landscape would contain local minima not relevant to the physical solution. 
(An identity transformation, for instance, satisfies other loss functions loss but does not correspond to the physical solutions of interest.)  
The total loss function is thus skewed toward a particular solution of the adversarial problem that also produces the relevant physics.
\begin{multline}\label{eqn:prediction_loss}
    \mathcal{L}_\text{pred}[(N,Z),r] = |\tilde{\zeta}_{(N,Z)+d} - T_d(\tilde{\zeta}_{(N,Z)})|\\
    + |\tilde{\zeta}_{(N,Z)} - T_{\bar{d}}(\tilde{\zeta}_{(N,Z)+d})|
\end{multline}

\item Cycle loss , Eq.~\ref{eqn:cycle_loss}, ensures that translators $T_d$ and $T_{\bar{d}}$ are asymptotically co-inverses, meaning changes to proton/neutron numbers must be reversible.
\begin{multline}\label{eqn:cycle_loss}
    \mathcal{L}_\text{cycle}[(N,Z),r] = |\tilde{\zeta}_{(N,Z)} - T_{\bar{d}}(T_d(\tilde{\zeta}_{(N,Z)}))|\\ 
    + |\tilde{\zeta}_{(N,Z)+d} - T_d(T_{\bar{d}}(\tilde{\zeta}_{(N,Z)+d}))|
\end{multline}

\item Discriminator loss, Eq.~\ref{eqn:discriminator_loss}, is low when the discriminator can accurately score training data as 1 (pass), and translated data and noise as 0 (fail).
\begin{multline}\label{eqn:discriminator_loss}
    \mathcal{L}_\text{disc}[(N,Z), r] = \text{BCE} \left[ D_r(\tilde{\zeta}_{(N,Z)},\tilde{\zeta}_{(N,Z)+d}) , 1 \right]\\
    + \text{BCE} \left[ D_r(\tilde{\zeta}_{(N,Z)}, T_d(\tilde{\zeta}_{(N,Z)})), 0 \right] \\
    + \text{BCE} \left[ D_r(T_{\bar{d}}(\tilde{\zeta}_{(N,Z)+d}),\tilde{\zeta}_{(N,Z)+d}), 0 \right]\\
    + \text{BCE} \left[ D_r(\text{noise},\text{noise}), 0 \right]\\
\end{multline}
where $\text{BCE}$ is binary crossentropy, an error measure often used for comparing probabilities.

\item Adversarial loss, Eq.~\ref{eqn:adversarial_loss}, is the converse of discriminator loss: it is low when the discriminator cannot distinguish between training data and translated data. In other words, training data and translated data are both scored 1 (pass), and random noise is scored 0 (fail).
\begin{multline}\label{eqn:adversarial_loss}
    \mathcal{L}_\text{adv}[(N,Z),r] = \text{BCE} \left[ D_r(\tilde{\zeta}_{(N,Z)}, T_d(\tilde{\zeta}_{(N,Z)})), 1 \right] \\
    + \text{BCE} \left[ D_r(T_{\bar{d}}(\tilde{\zeta}_{(N,Z)+d}),\tilde{\zeta}_{(N,Z)+d}), 1 \right]\\
    + \text{BCE} \left[ D_r(T_{\bar{d}}(\tilde{\zeta}_{(N,Z)+d}),T_{d}(\tilde{\zeta}_{(N,Z)})), 1 \right]\\
    + \text{BCE} \left[ D_r(\text{noise},\text{noise}), 0 \right]\\
\end{multline}

\end{itemize}

The total cycleGAN loss function is 
\begin{equation}\label{eqn:total_loss}
    \mathcal{L}_\text{GAN} = \lambda_\text{pred} \mathcal{L}_\text{pred}
    + \mathcal{L}_\text{cycle}
    + \lambda_\text{adv} \mathcal{L}_\text{adv} 
    + \lambda_\text{disc} \mathcal{L}_\text{disc}
\end{equation}
The coefficients $\lambda$ are tunable hyperparameters chosen by trial-and-error.
We found that introducing $\lambda_\text{adv}=5-10$ improved translator convergence; in practice even a relatively small discriminator network can overpower the tranlators. 
The prediction loss coefficient $\lambda_\text{pred}$ is set larger (around 10) at the beginning of training, then scheduled to decrease to $\leq1$ as the networks converge. This brings the system to the relevant region of parameter space early in training without overfitting. 

The GAN over/underfitting problem \cite{yazici2020empirical} is subtle and challenging, in part because it is quite different from fitting well-understood regression models. 
We try to control over/underfitting mainly through controlling model complexity and implementing regularization.
Several network configurations were tested, depths and widths, for both translators and discriminators; by far the best results came from our largest configuration, with translators having 4 hidden layers of size 512 and discriminators having 4 hidden layers of size 68.
Regularization of the networks is implemented in two ways: dropout and label noise. 
Dropout \cite{srivastava2014dropout} is used in both the generators and the discriminator; during training, upon each iteration, some percentage of weights are temporarily set to zero. 
This makes network predictions more robust since no single neuron or set of neurons can be relied on all of the time, and making it much less likely that the network store any static information.
We use a dropout probability of 50\% on the inner hidden layers of the translators and 25\% for the discriminators. 
Label noise \cite{jenni2019stabilizing,wiatrak2019stabilizing} is a regularization tool that works on the discriminator; rather than using labels 0 or 1 for all training data, a small random $\epsilon$ is chosen and added/subtracted to the label accordingly. That is, upon every iteration we choose $\epsilon \sim \mathcal{N}(0,\sigma_\epsilon^2)$ and then replace $0 \rightarrow 0 + |\epsilon|$ and $ 1 \rightarrow 1 - |\epsilon|$. The standard deviation $\sigma_\epsilon$ is 0.1 in the present application (this value was not arrived at via any optimization, and in general, such an exploration may be warranted). 
This has the effect of smoothing the discriminator predictions around the limits, keeping the discriminator from assigning probabilities ``overconfidently'', so to speak. 
Furthermore, we assisted the adversarial nature of the model slightly by introducing a ``freezing'' value for each loss term; when the discriminator loss drops below a freezing point, its weights are fixed to allow the translator to catch up. The translators similarly have their own freezing value.  

Development and training were carried out on heterogeneous architecture at Lawrence Livermore National Laboratory with IBM POWER9 chips: per training session, the CPU creates the relevant data environment, including the cycleGAN model with appropriate hyperparameters, and sends it to the on-board GPU for weight optimization. 
Optimization was performed using ADAM \cite{adam_kingma} with a moving average wrapper \cite{gan_averaging_yazici}. 
ADAM already includes a momentum contribution, but the moving average means the weights are even less sensitive to small changes in the loss function, which is very helpful because gradients of adversarial losses can be very noisy.

\section{\label{sec:predictions}Predictions \& validation}

Accuracy of predictions is judged using using validation data, as in a typical regression problem: we assume that when errors on validation data are approximately equal to those on training data, locally, the prediction quality is good. 
We prepared two data sets for experimentation: dataset $\mathcal{A}$ has a subset of 20 cross sections distributed along the length of the chart held out for validation, and dataset $\mathcal{B}$ has a 3x3 region with $^{170}$Yb at the center, nine cross sections total, held out as a validation set.
Two copies of the entire model were trained on the respective datasets.

\subsection{Predicting direct neighbors}

\begin{figure*}
\includegraphics[scale=0.5]{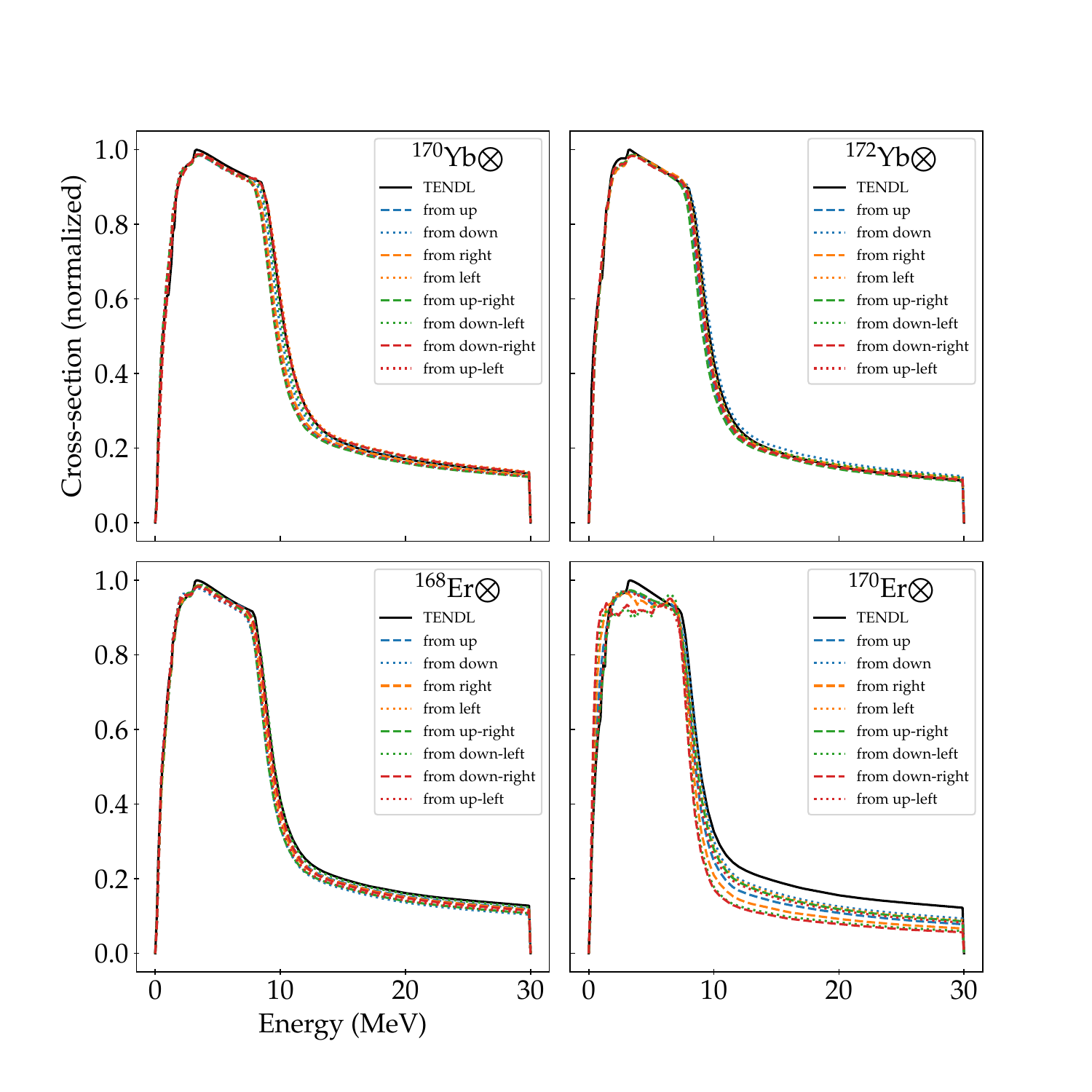}
\caption{Local predictions from data set $\mathcal{A}$.  In this example, all shown cross sections were excluded from the training data (marked with $\otimes$), and some (but not all) neighbors were included. This corresponds to the region inside the $3\times3$ green square in Fig.~\ref{fig:chart_heatmap_leavout3x3}, which is centered on $^{170}$Yb. 
The changes between curves are very small, and predictions are accurate.}
\label{fig:local_predictions_LO3x3_1}
\end{figure*}

\begin{figure*}
\includegraphics[scale=0.5]{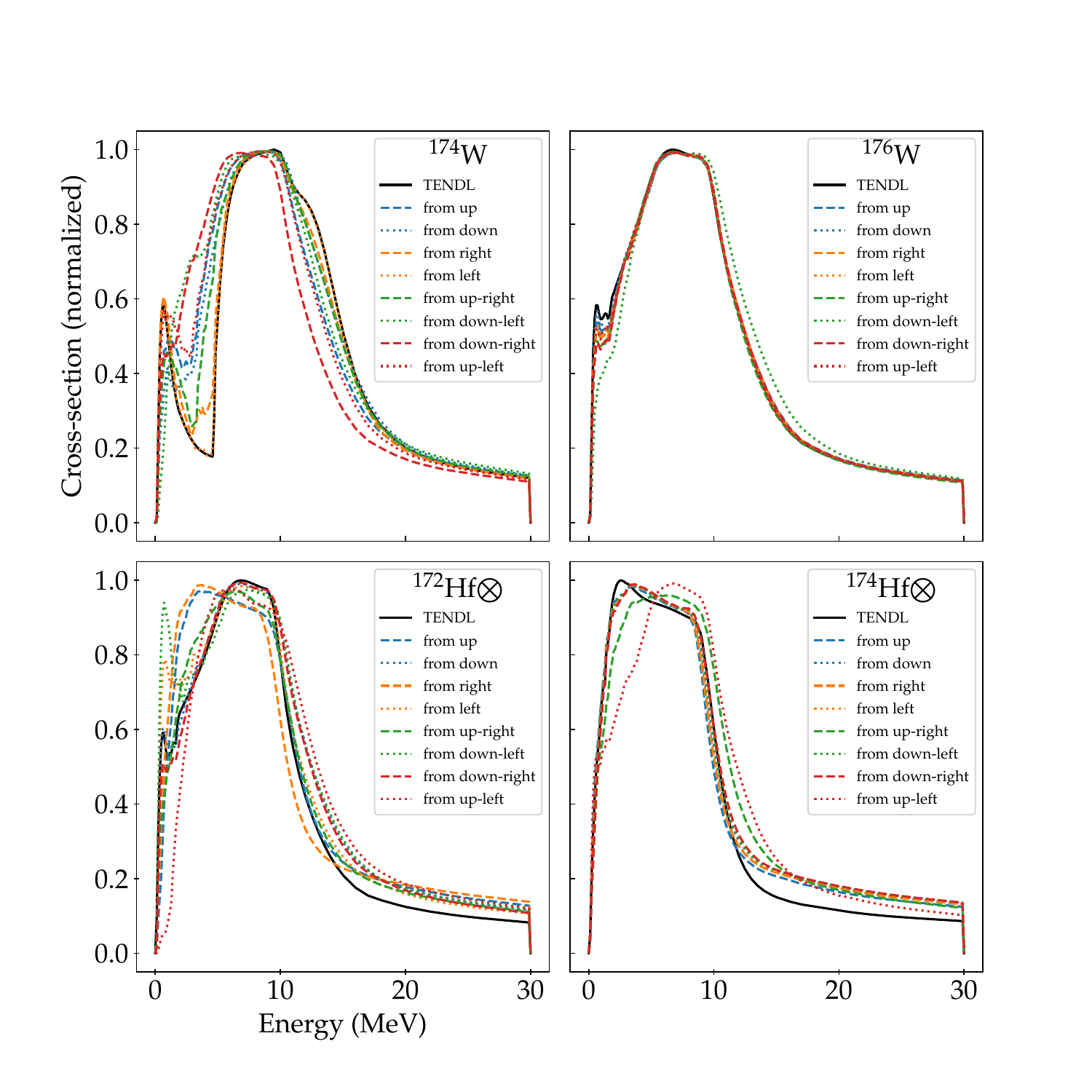}
\caption{Local predictions from data set $\mathcal{A}$. The bottom two cross sections were excluded from the training data (marked with $\otimes$), as were several others below those. 
The predictions for $^{174}$W show significant disagreement, but this is present even when no validation data is held out, indicating that the low-energy structure represents a significant break from general trends.}
\label{fig:local_predictions_LO3x3_2}
\end{figure*}

\begin{figure*}
\includegraphics[scale=0.5]{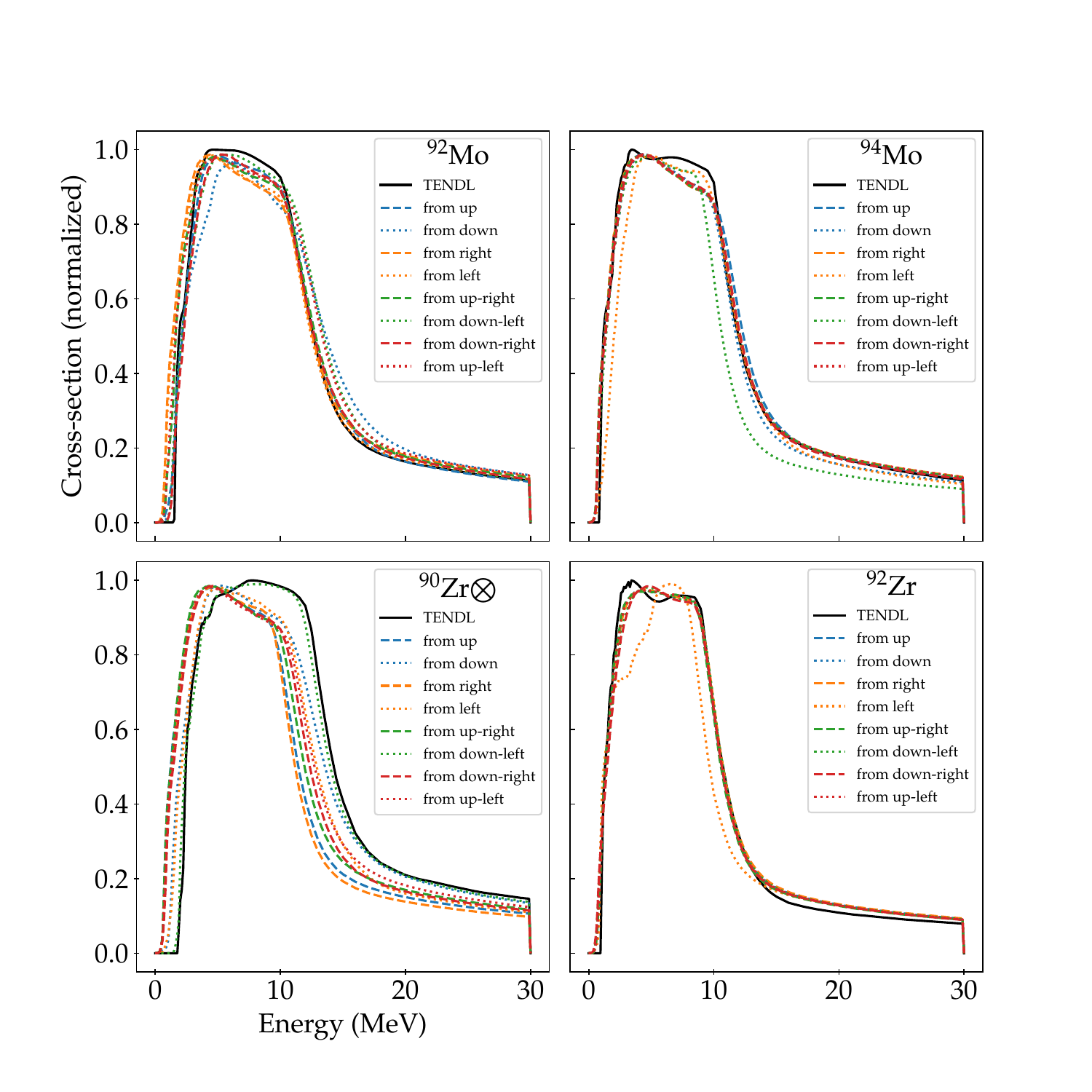}
\caption{Local predictions from data set $\mathcal{B}$. In this example, the $^{90}$Zr cross section was excluded from the training data, and almost all translators predict a slightly different curve that looks more like those in the region.}
\label{fig:local_predictions_DC20_1}
\end{figure*}

\begin{figure*}
\includegraphics[scale=0.5]{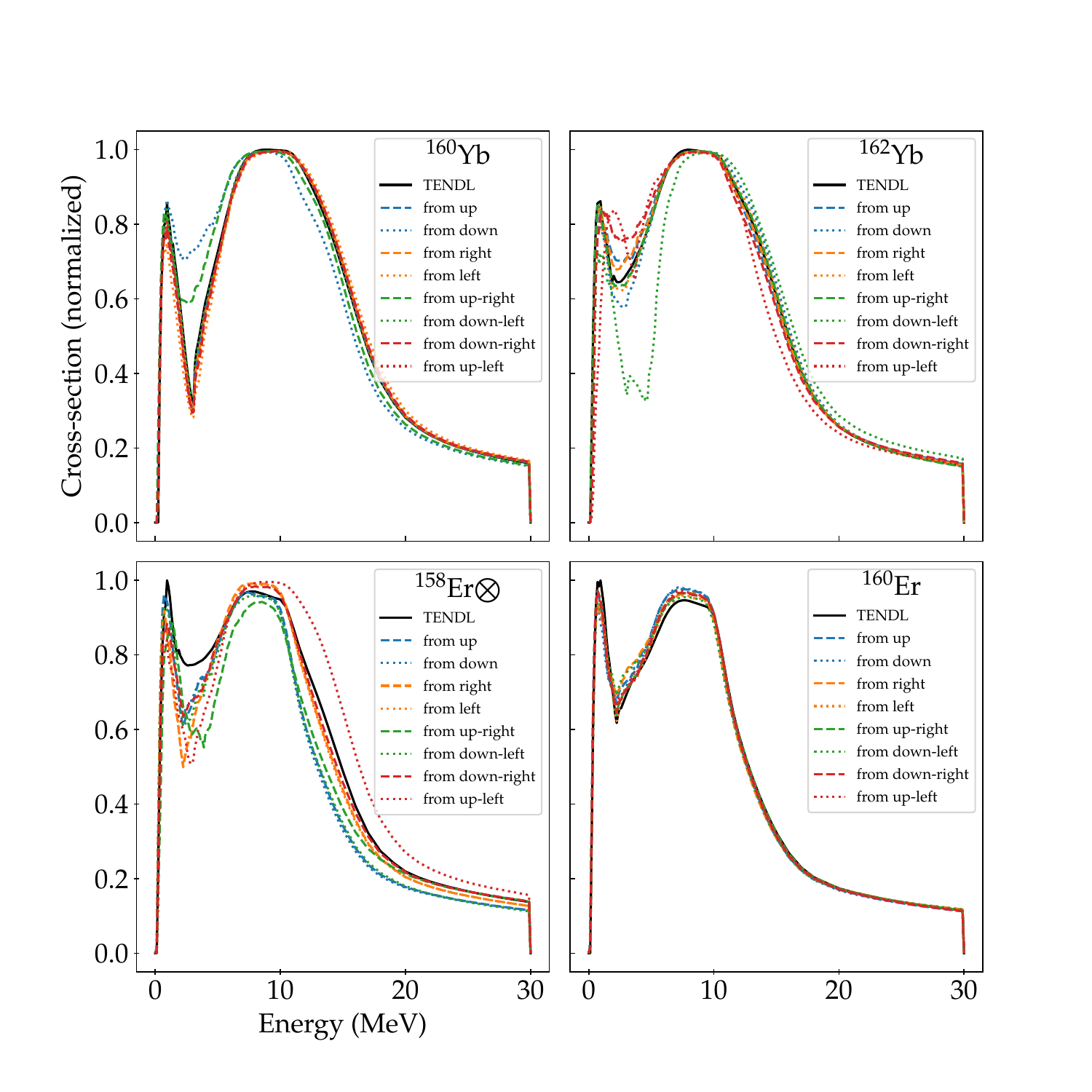}
\caption{Local predictions from data set $\mathcal{B}$. The $^{158}$Er cross section was excluded from the training data. Interestingly, the translators correctly predict the general shape of the withheld curve, but overestimate the low-energy cut in cross section typical of nearby nuclides. }
\label{fig:local_predictions_DC20_2}
\end{figure*}

The simplest use of the model is to predict cross sections of direct neighbors. 
Figures \ref{fig:local_predictions_LO3x3_1} and \ref{fig:local_predictions_LO3x3_2} show $2 \times 2$ regions of local predictions using dataset $\mathcal{A}$. 
Figures \ref{fig:local_predictions_DC20_1} and \ref{fig:local_predictions_DC20_2} show $2 \times 2$ regions of local predictions using dataset $\mathcal{B}$. 
In each subplot, the solid black line is the TENDL cross section curve, and each colored dashed line is the prediction from a direct neighbor (e.g., a curve labeled ``from up-left'' is the prediction of $\sigma_{(N,Z)}$ made by evaluating $T_\searrow$ on $\sigma_{(N-2,Z+2)}$). 
Nuclei excluded from the training set are marked with $\otimes$ in those figures.
In that case, the model did not receive any information about that nuclide.
We have included many more plots in the supplemental material.

\begin{figure*}
\includegraphics[width=\textwidth]{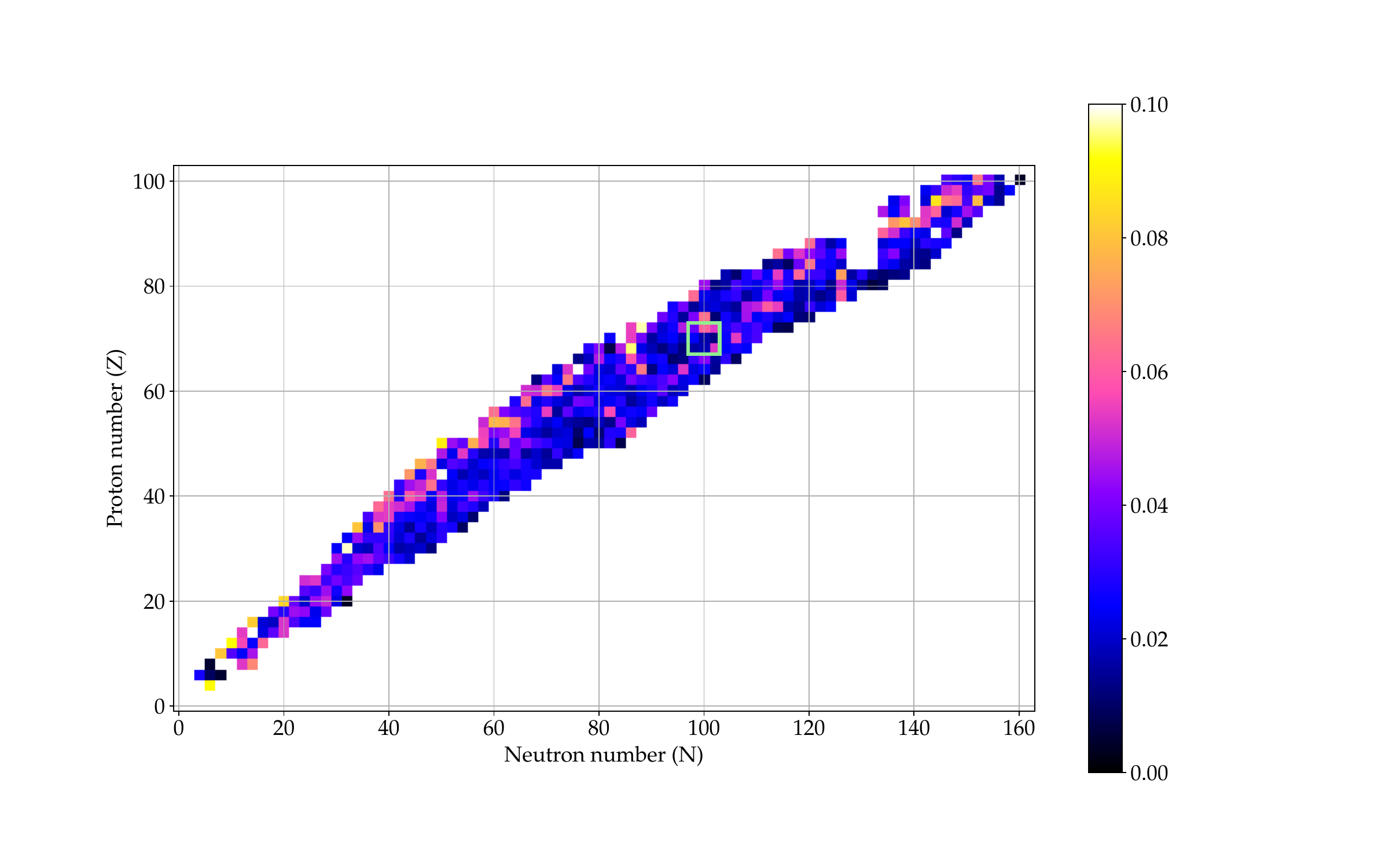}
\caption{Average error from immediate neighbors, trained on data without 3x3 region around $^{170}$Yb (outlined in green). We find that, when comparing to a model trained on all data, the errors look very similar. The nuclides around the top of the withheld region exhibit some unique changes that prove challenging regardless of what data is used for traning.}
\label{fig:chart_heatmap_leavout3x3}
\end{figure*}

\begin{figure*}
\includegraphics[width=\textwidth]{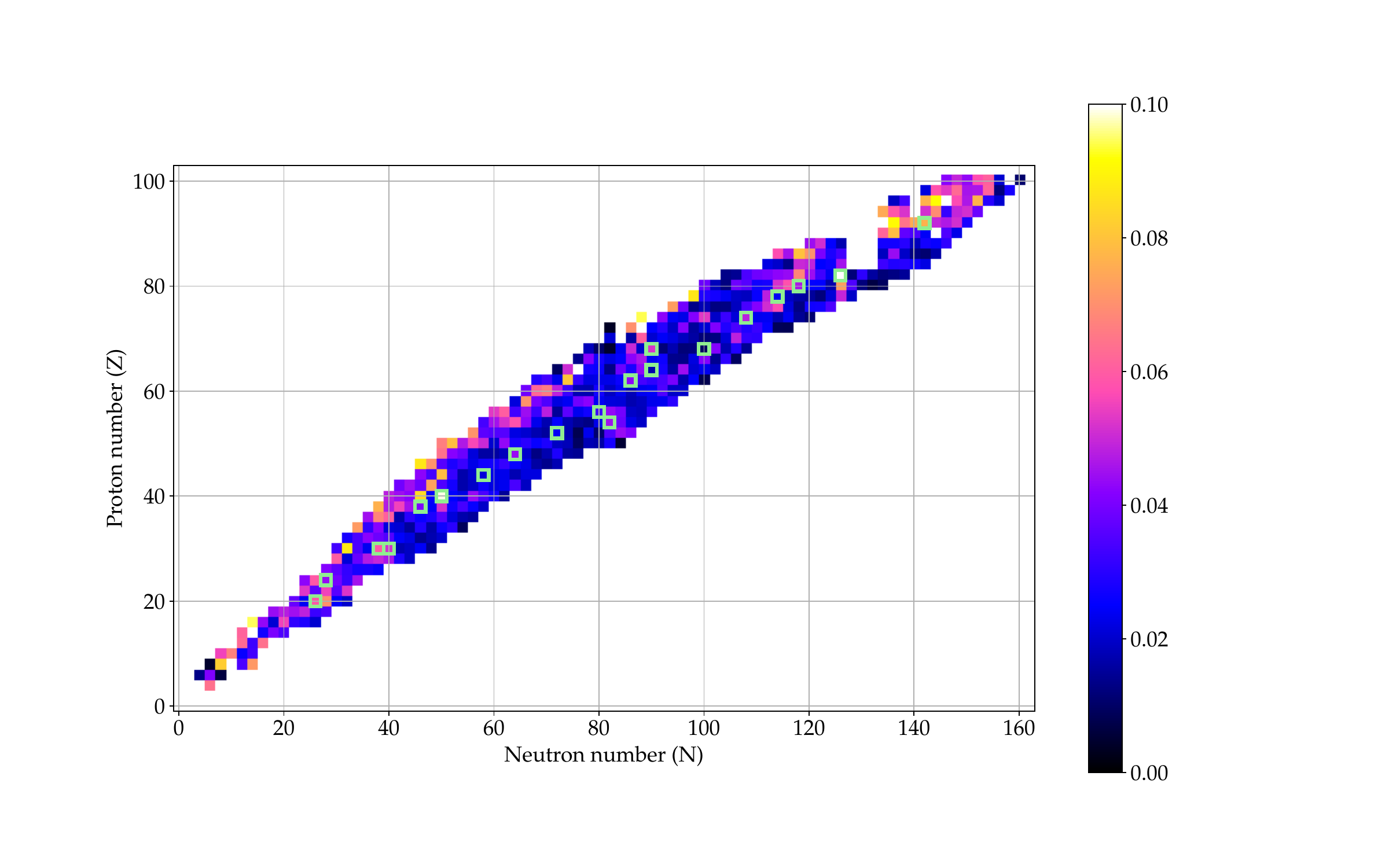}
\caption{Average error from immediate neighbors, trained on data without a scattered subset of 20 nuclides (outlined in green). This model had been trained for less time than that shown in Fig.~\ref{fig:chart_heatmap_leavout3x3}, hence the larger errors for some nuclei. Once again, we see that on average the model is robust to dropping data points, but only if those curves do not break drastically from local trends. In those cases, the errors are indeed larger.}
\label{fig:chart_heatmap_dropcenter20}
\end{figure*}

The error heatmaps for datasets $\mathcal{A}$ and $\mathcal{B}$, shown in Figs.~\ref{fig:chart_heatmap_dropcenter20} and \ref{fig:chart_heatmap_leavout3x3} respectively, show a global picture of local predictions, with average MAE values represented as colors, and each tile is a single cross section. 
Validation data are outlined in light green. 
In the case of both datasets, we find that the distribution of errors in the regions of validation data match the distributions of errors when no data was withheld.


\subsection{Linked translations}

\begin{figure*}
\begin{subfigure}{\textwidth}
\includegraphics[scale=0.5]{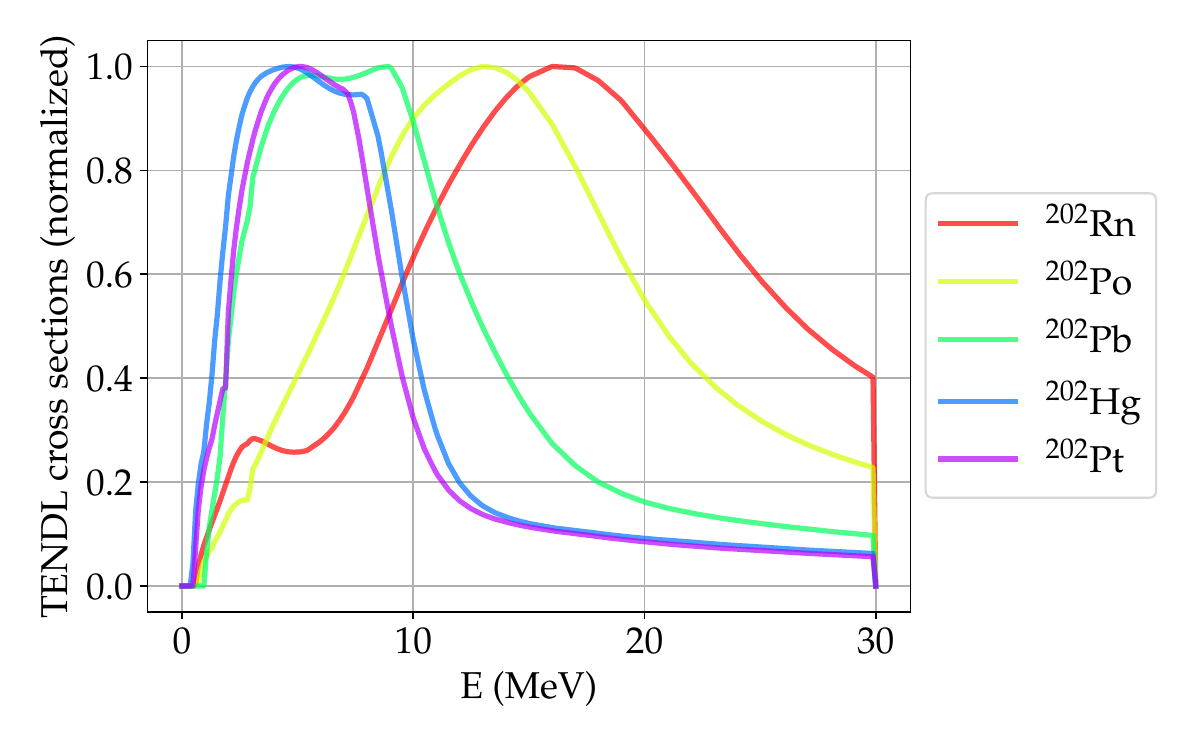}
\caption{TENDL cross sections }
\end{subfigure}
\bigskip
\begin{subfigure}{\textwidth}
\includegraphics[scale=0.5]{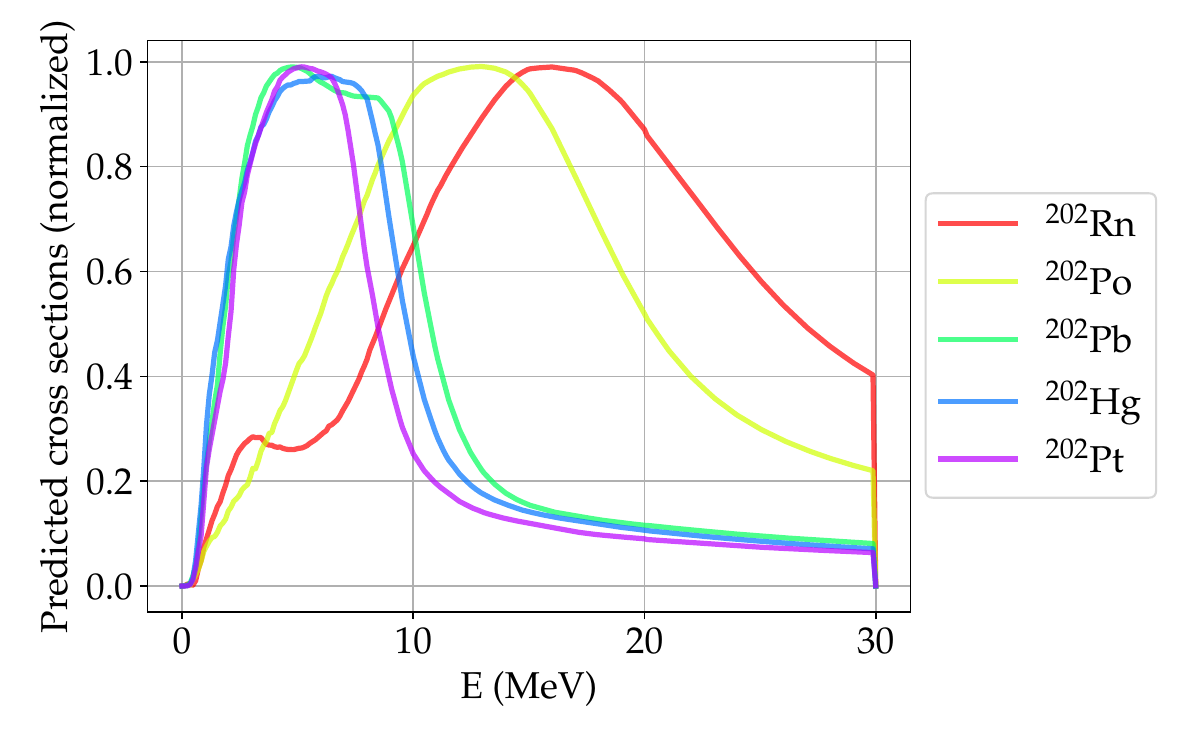}
\caption{GAN predictions }
\end{subfigure}
\caption{Cross sections in the $(N,Z) \rightarrow (N+2,Z-2)$ direction ($\searrow$), beginning at $^{202}$Rn for 5 steps. In this case, the GAN reproduces the larger trends well, and only smaller features are lost by the end of the interpolation (purple curve).}
\label{fig:ray_extrap_1}
\end{figure*}

\begin{figure*}
\begin{subfigure}{\textwidth}
\includegraphics[scale=0.5]{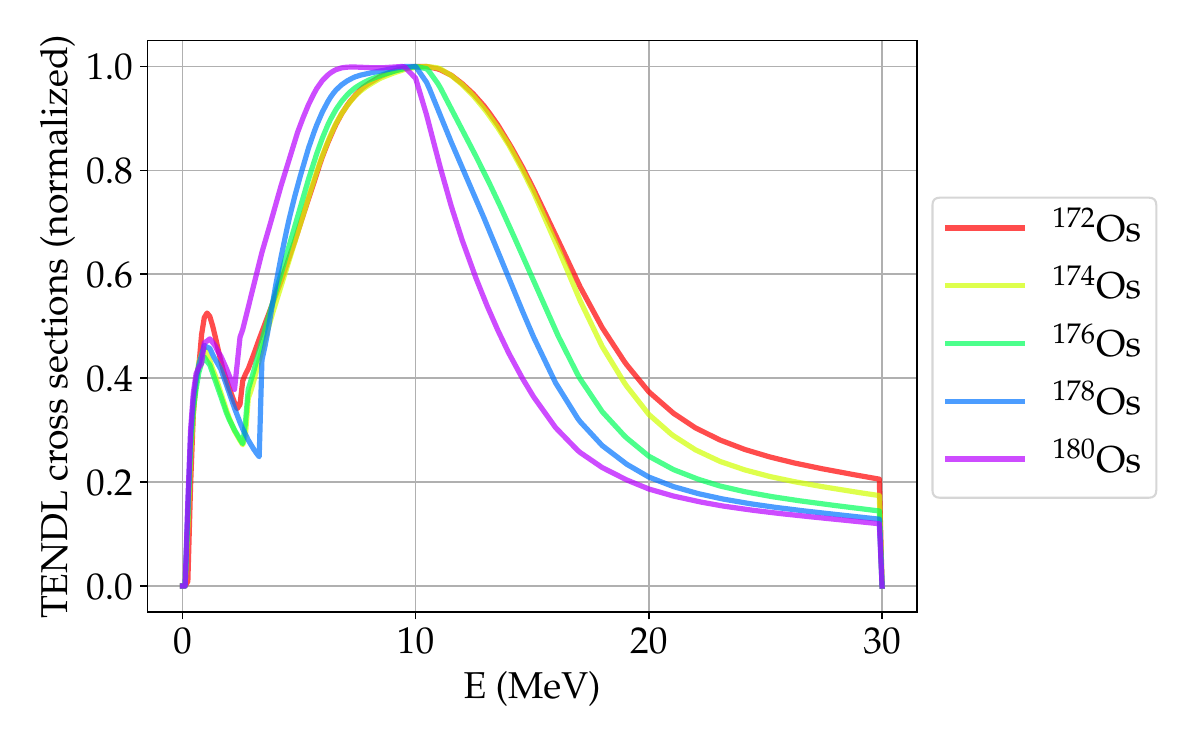}
\caption{TENDL cross sections }
\end{subfigure}
\bigskip
\begin{subfigure}{\textwidth}
\includegraphics[scale=0.5]{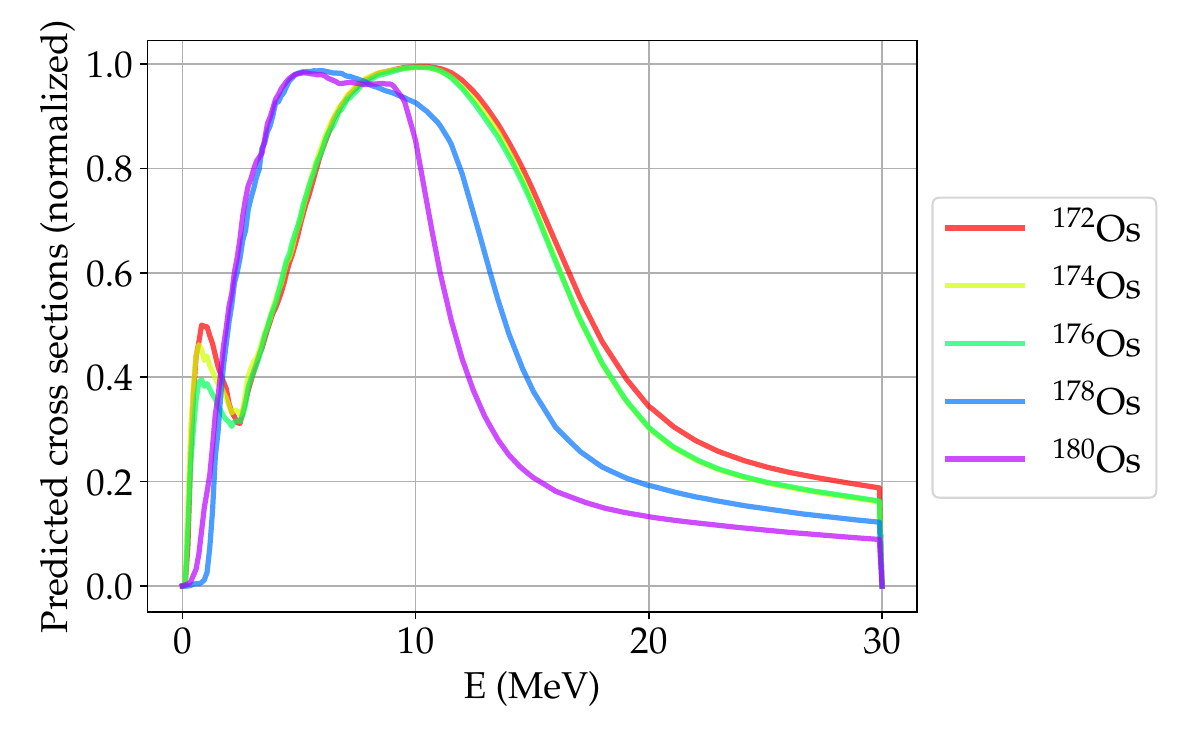}
\caption{GAN predictions }
\end{subfigure}
\caption{Cross sections in the $(N,Z) \rightarrow (N+2,Z)$ direction ($\rightarrow$), beginning at $^{172}$Os for 5 steps. The GAN makes good predictions for the first three steps but changes at $^{178}$Os, beyond which it no longer reproduces the low-energy structure or long tail.}
\label{fig:ray_extrap_2}
\end{figure*}

\begin{figure*}
\begin{subfigure}{\textwidth}
\includegraphics[scale=0.5]{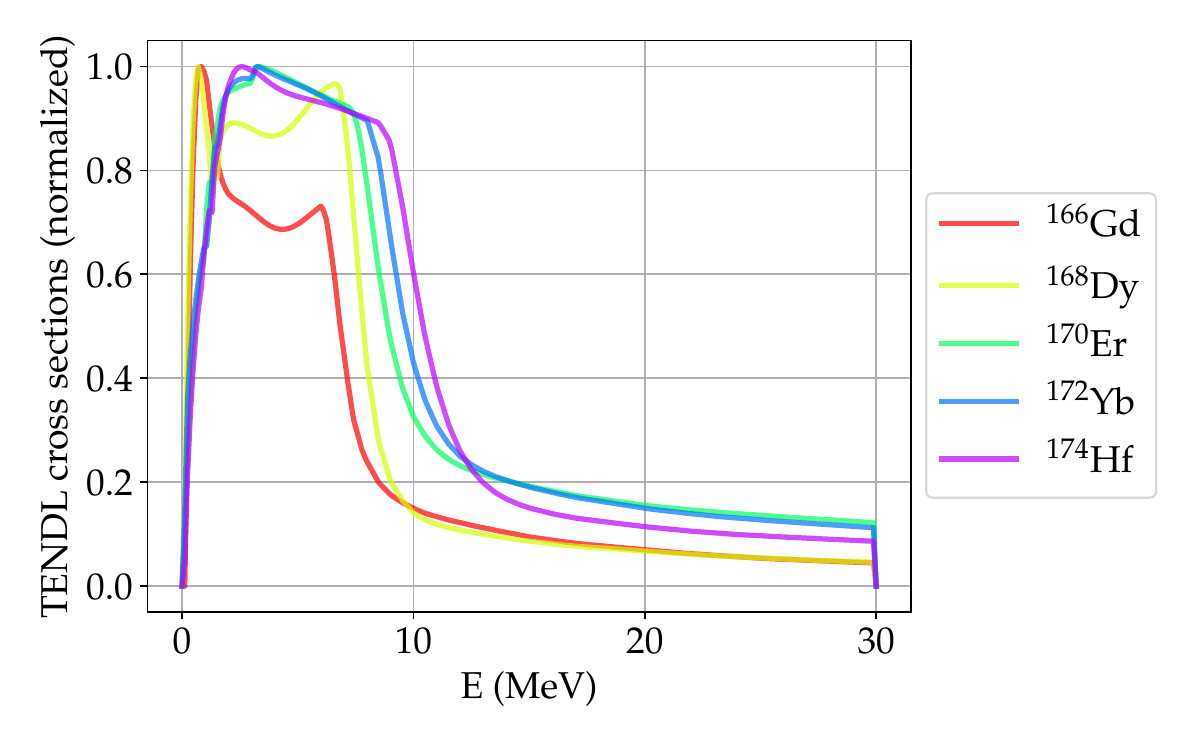}
\caption{TENDL cross sections }
\end{subfigure}
\bigskip
\begin{subfigure}{\textwidth}
\includegraphics[scale=0.5]{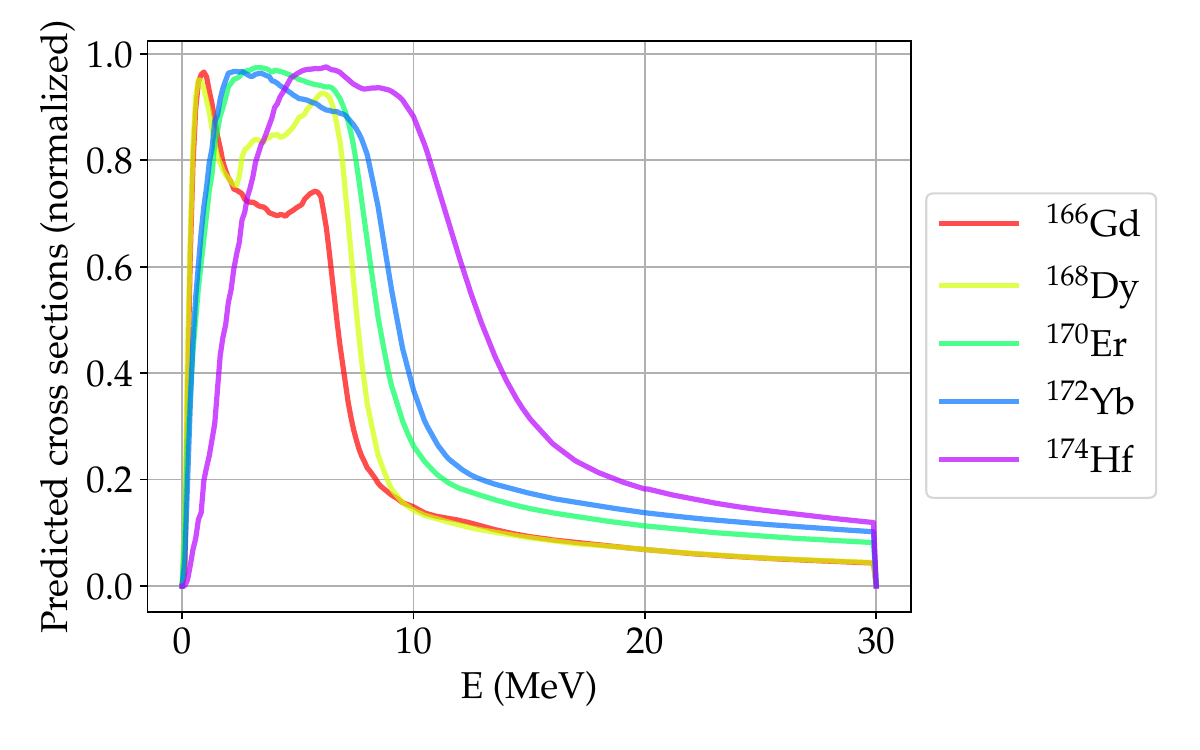}
\caption{GAN predictions }
\end{subfigure}
\caption{Cross sections in the $(N,Z) \rightarrow (N,Z+2)$ direction ($\uparrow$), beginning at $^{166}$Gd for 5 steps.  This sequence is reproduced nicely by the model until the fifth step to $^{174}$Hf, which shows noticeable errors.}
\label{fig:ray_extrap_3}
\end{figure*}

\begin{figure*}
\begin{subfigure}{\textwidth}
\includegraphics[scale=0.5]{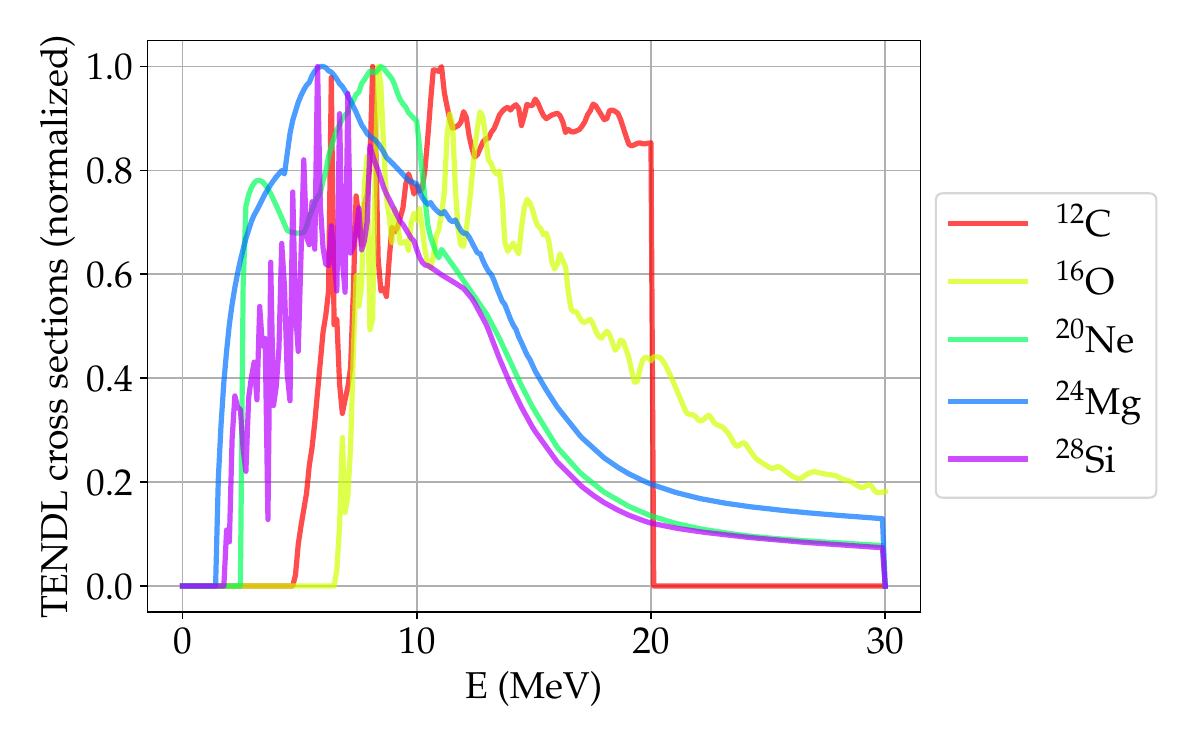}
\caption{TENDL cross sections }
\end{subfigure}
\bigskip
\begin{subfigure}{\textwidth}
\includegraphics[scale=0.5]{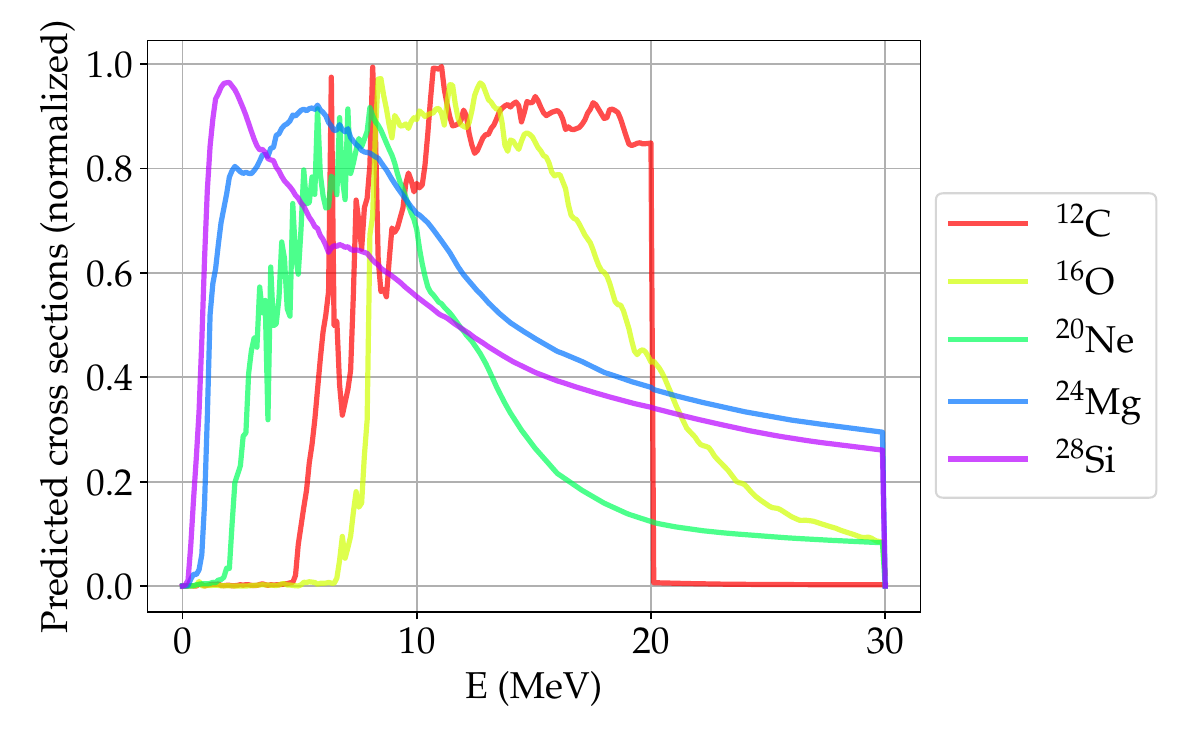}
\caption{GAN predictions }
\end{subfigure}
\caption{Cross sections in the $(N,Z) \rightarrow (N+2,Z+2)$ direction ($\nearrow$), beginning at $^{12}$C for 5 steps. Here we see the GAN reconstructing the $^{12}$C curve (red) very accurately, but beyond that, the predictions show significant errors. These inaccuracies are due to the fine-tuning of cross section data for this region of the chart (note the high-frequency fluctuations), but also, the trends present in $Z_\neswarrow$ are arguably more complex than other subsets (due to changes in both $Z$ and $A$). In that sense, these predictions are the most challenging, and the large inaccuracy shown here is expected.}
\label{fig:ray_extrap_4}
\end{figure*}

We can link translations in paths across the chart and predict curves for nuclides beyond direct neighbors.
This process is not restricted to the region of training data. 
The accuracy and stability of predictions appear to be dependent on the complexity of local trends around where the predictions are made. 
Each chain of predictions is stable for a number of steps, empirically around 5-6 on average, but can be longer or shorter depending on trends locally present and the introduction of anomalous data.
Some examples of interpolation are given in Figs.~\ref{fig:ray_extrap_1}, \ref{fig:ray_extrap_2}, \ref{fig:ray_extrap_3}, and \ref{fig:ray_extrap_4}. 
These examples show a list of TENDL cross sections in one plot and a list of extrapolated GAN predictions in the other. 
The first predicted curve $\sigma'_{(N,Z)}$ is simply the first cross section $\sigma_{(N,Z)}$ encoded and decoded once, as in 
\begin{equation}
    \sigma'_{(N,Z)} = E^+(E(\sigma_{(N,Z)})).
\end{equation} 
The second, $\sigma'_{(N,Z)+d}$, is the first cross section encoded, translated once, and decoded, as in
\begin{equation}
 \sigma'_{(N,Z)+d} = E^+(T_d(E(\sigma_{(N,Z)}))) ,
\end{equation}
where $d$ is the direction of translation.  
This may be repeated for $n$ steps: $\sigma'_{(N,Z)+nd} = E^+(T^n_d(E(\sigma_{(N,Z)})))$.

Furthermore, translations in the path need not be restricted to one rule; relaxing this constraint leads to the ensemble capabilities discussed in the next section. 
Consider a convolution of the form

\begin{equation}
    \sigma'_{(N',Z')} = E^+ \left( \prod_{d\in\mathcal{D}} T_d(E(\sigma_{(N,Z)}) \right),
    \label{eq:conv_product}
\end{equation}

where $\mathcal{D}$ is a sequence of directions. This convolution is tantamount to moving along a path on the chart of nuclides, beginning at $(N,Z)$, following steps in the sequence $\mathcal{D}$, and terminating at $(N',Z')$.

\subsection{Ensemble predictions}

Leveraging the prediction capability, we can perform \textbf{ensemble predictions} for cross sections beyond the training set as well.  Ensemble methods can be used in many different machine learning problems and are useful because the model(s) produce a set of predictions rather than just one, and thus we have a built-in measure of variability. In our GAN, this consists of computing many different paths across the chart and predicting the cross section for a common final nuclide. 
Neural network models are commonly understood to have difficulty with extrapolations, at least without the proper modifications. R
egularization, which is basically an additional constraint, can help us produce better extrapolations. In particular, distribution learning and constraints placed on latent space distributions of data can be helpful. 
The VAE, for instance, limits the distribution of latent variables using variational Bayes, and this results in a fantastically smooth latent representation (see \cite{kingma2013_vae} for more). 
Our working hypothesis is that distribution learning may be leveraged to achieve good extrapolation behavior; this certainly helps achieve good results within this work, but proof of this concept does not, to our knowledge, exist.

Ensemble predictions are formed by a set of individual prediction chains. We first designate a target nuclide and a region of the chart for the starting points. Then, we compute all possible paths which begin within our region of interest and terminate on the target. 
An added constraint is that the paths do not ``back-track''; that is, within each path, the distance to the target nuclide is monotonically decreasing with each step. This property ensures that we are not including undue errors in the final ensemble. 
In this way, the calculation in Eq.~\ref{eq:conv_product} is done for many different paths $\mathcal{D}$, which constitute a sampling of $P(\sigma'|\mathcal{D})$, the distribution of ensemble predictions.  

\begin{figure*}
\begin{subfigure}{\textwidth}
\includegraphics[scale=0.4]{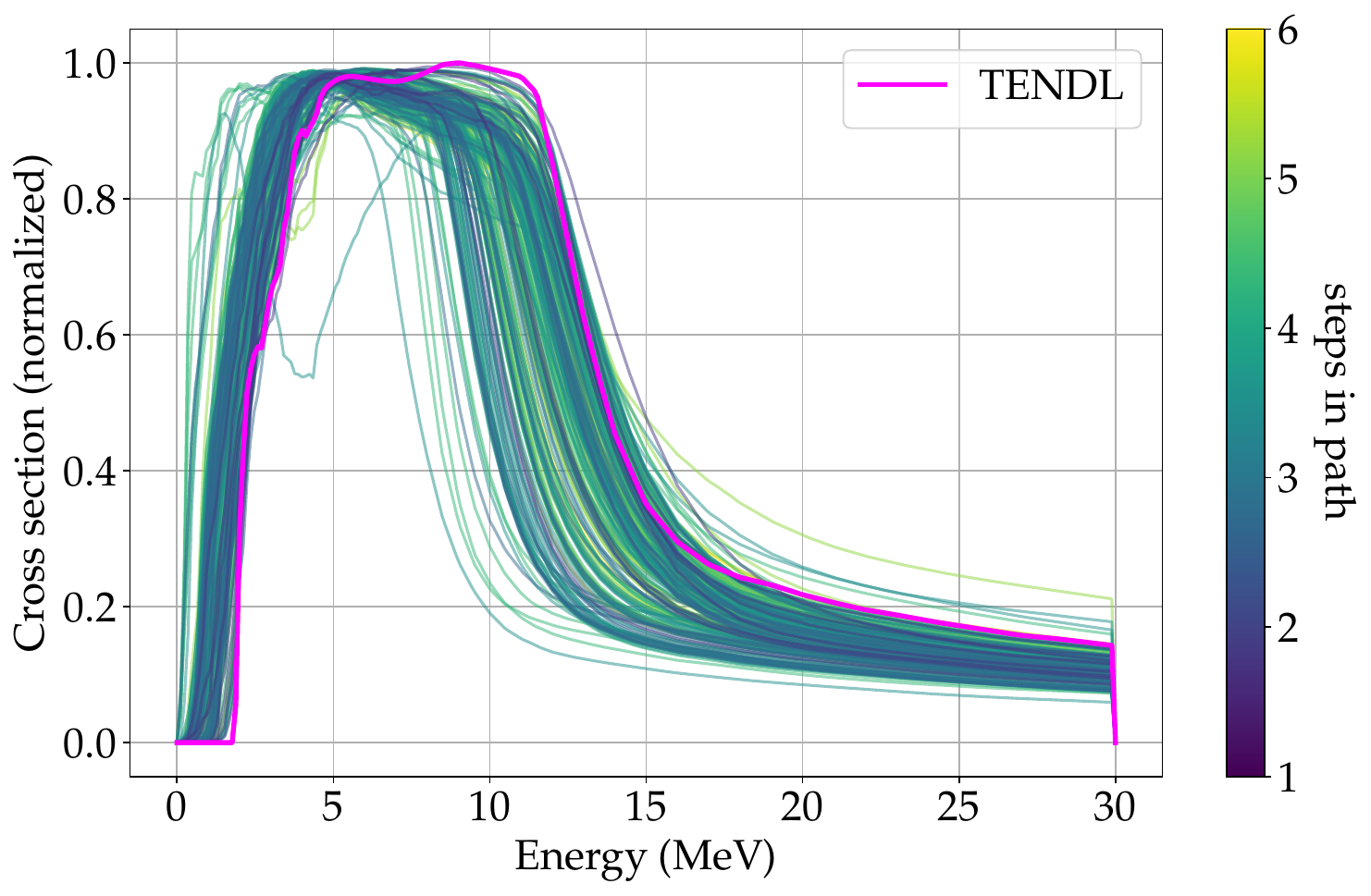}
\caption{Ensemble prediction of $^{88}$Sr with all 273 predicted curves shown. GAN predictions are shown in viridis (blue-green-yellow) where the color corresponds to the number of linked translations. }
\end{subfigure}
\bigskip
\begin{subfigure}{\textwidth}
\includegraphics[scale=0.4]{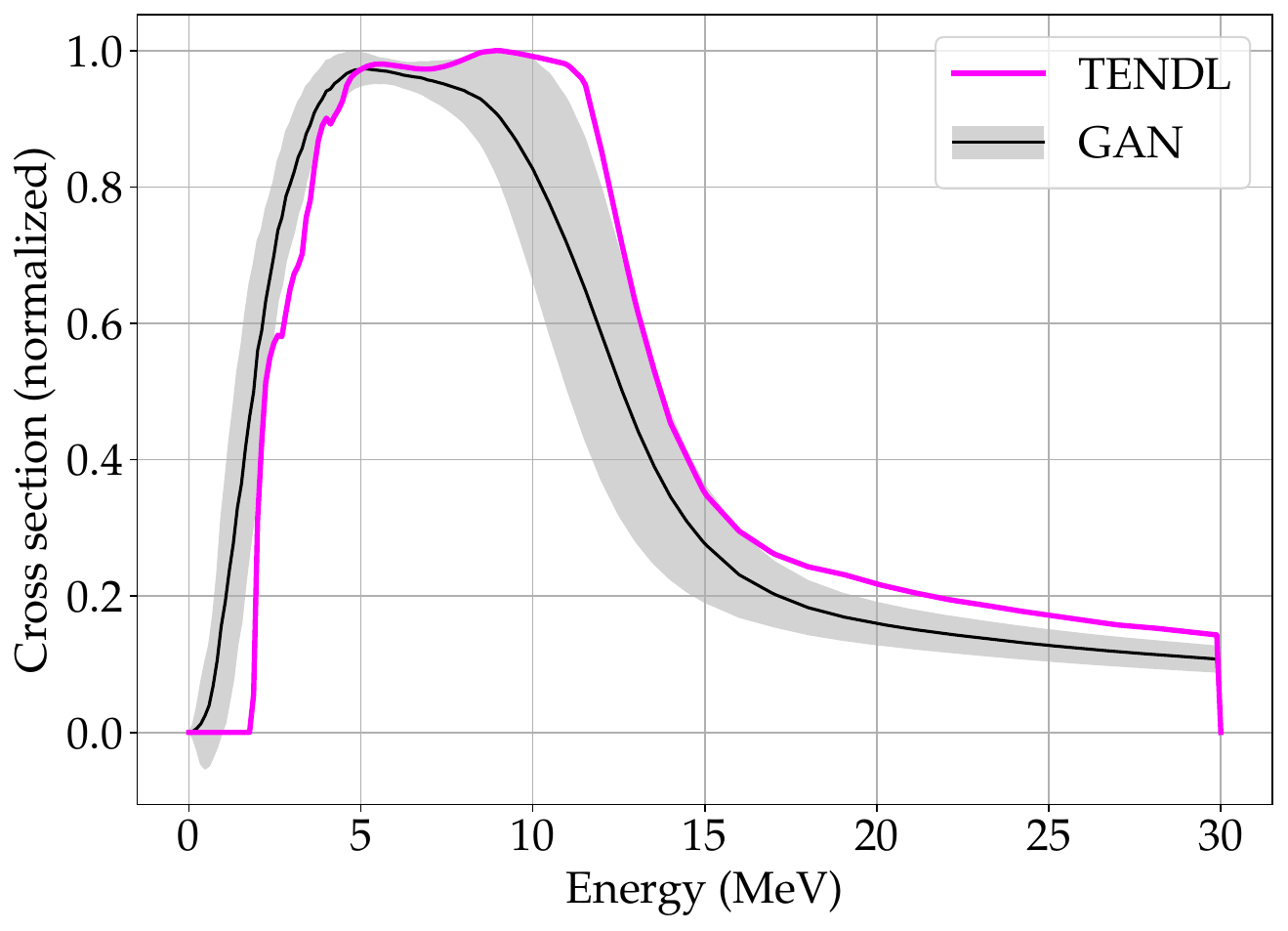}
\caption{Ensemble prediction of $^{88}$Sr represented by a median and standard deviation band, which is computed using a weighted average of curves in the ensemble.}
\end{subfigure}
\caption{Ensemble prediction of the (normalized) cross section of Strontium-88, using all possible paths within a bounded local region of the chart of nuclides. 
Almost all curves in our ensemble are reasonable in shape, but many show the reaction turning off at lower energies than observed.}
\label{fig:ml:ensemble_prediction_88Sr}
\end{figure*}

\begin{figure*}
\begin{subfigure}{\textwidth}
\includegraphics[scale=0.4]{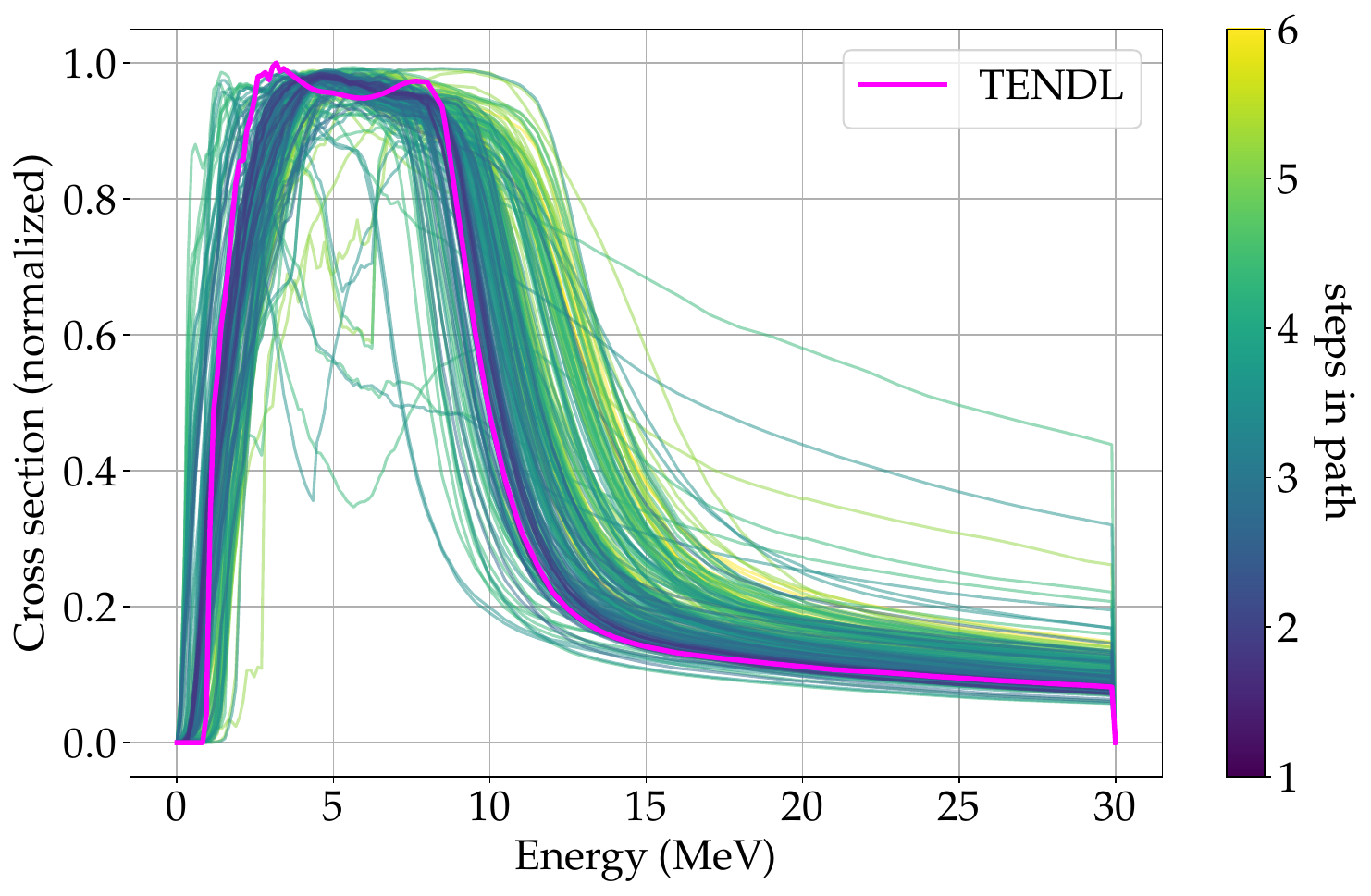}
\caption{Ensemble prediction of $^{94}$Zr with all 260 predicted curves shown. GAN predictions are shown in viridis (blue-green-yellow) where the color corresponds to the number of linked translations.}
\end{subfigure}
\bigskip
\begin{subfigure}{\textwidth}
\includegraphics[scale=0.4]{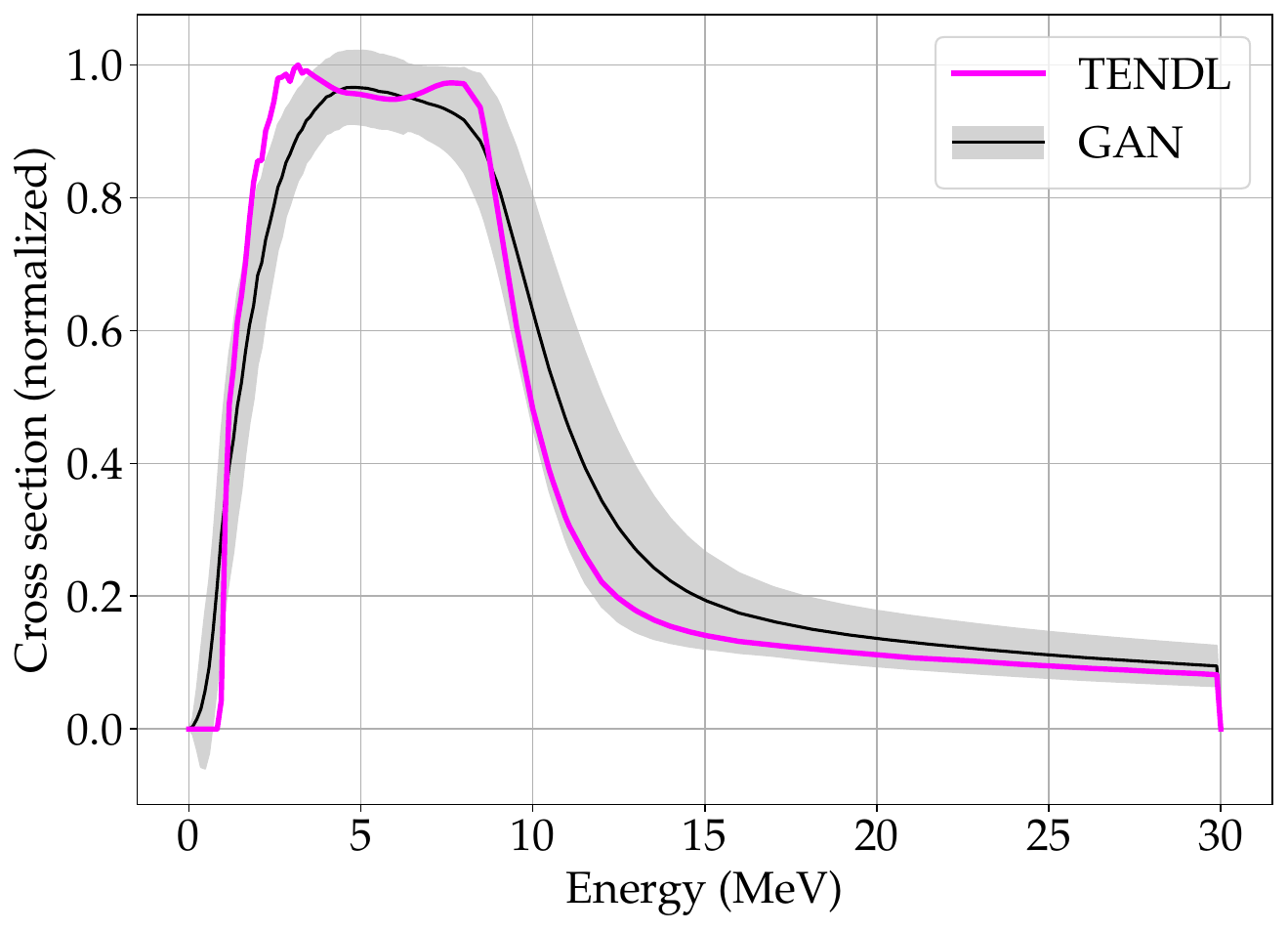}
\caption{Ensemble prediction of $^{94}$Zr represented by a median and standard deviation band, which is computed using a weighted average of curves in the ensemble.}
\end{subfigure}
\caption{Ensemble prediction of the (normalized) cross section of Zirconium-94, using all possible paths within a bounded local region of the chart of nuclides.
This example is included to illustrate the benefit of averaging over the ensemble: some individual predictions have large errors but the ensemble median predicts the cross section well.}
\label{fig:ml:ensemble_prediction_94Zr}
\end{figure*}

\begin{figure*}
\begin{subfigure}{\textwidth}
\includegraphics[scale=0.4]{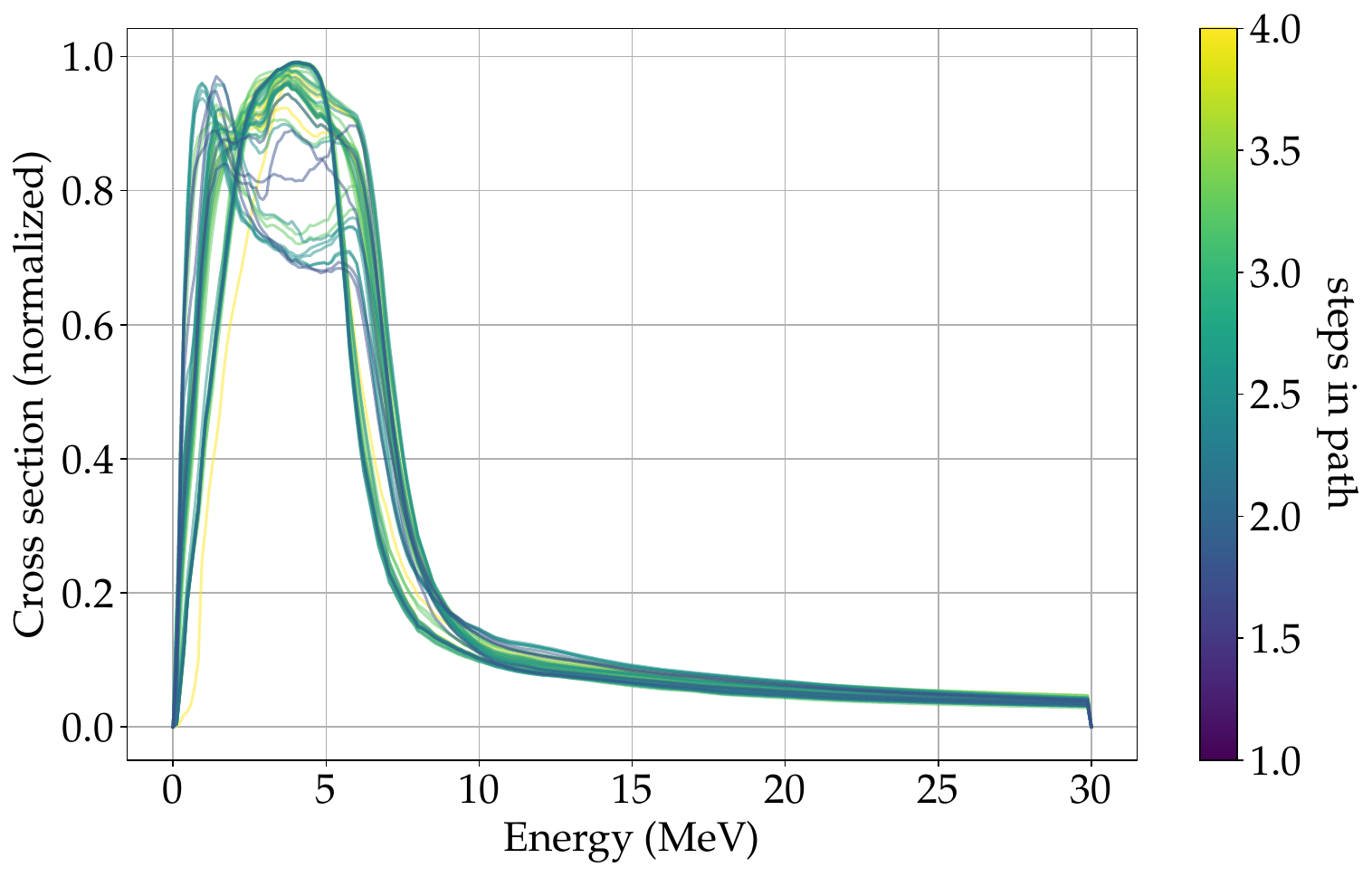}
\caption{Ensemble prediction of $^{158}$Ce with all 57 predicted curves shown. GAN predictions are shown in viridis (blue-green-yellow) where the color corresponds to the number of linked translations.  }
\end{subfigure}
\bigskip
\begin{subfigure}{\textwidth}
\includegraphics[scale=0.4]{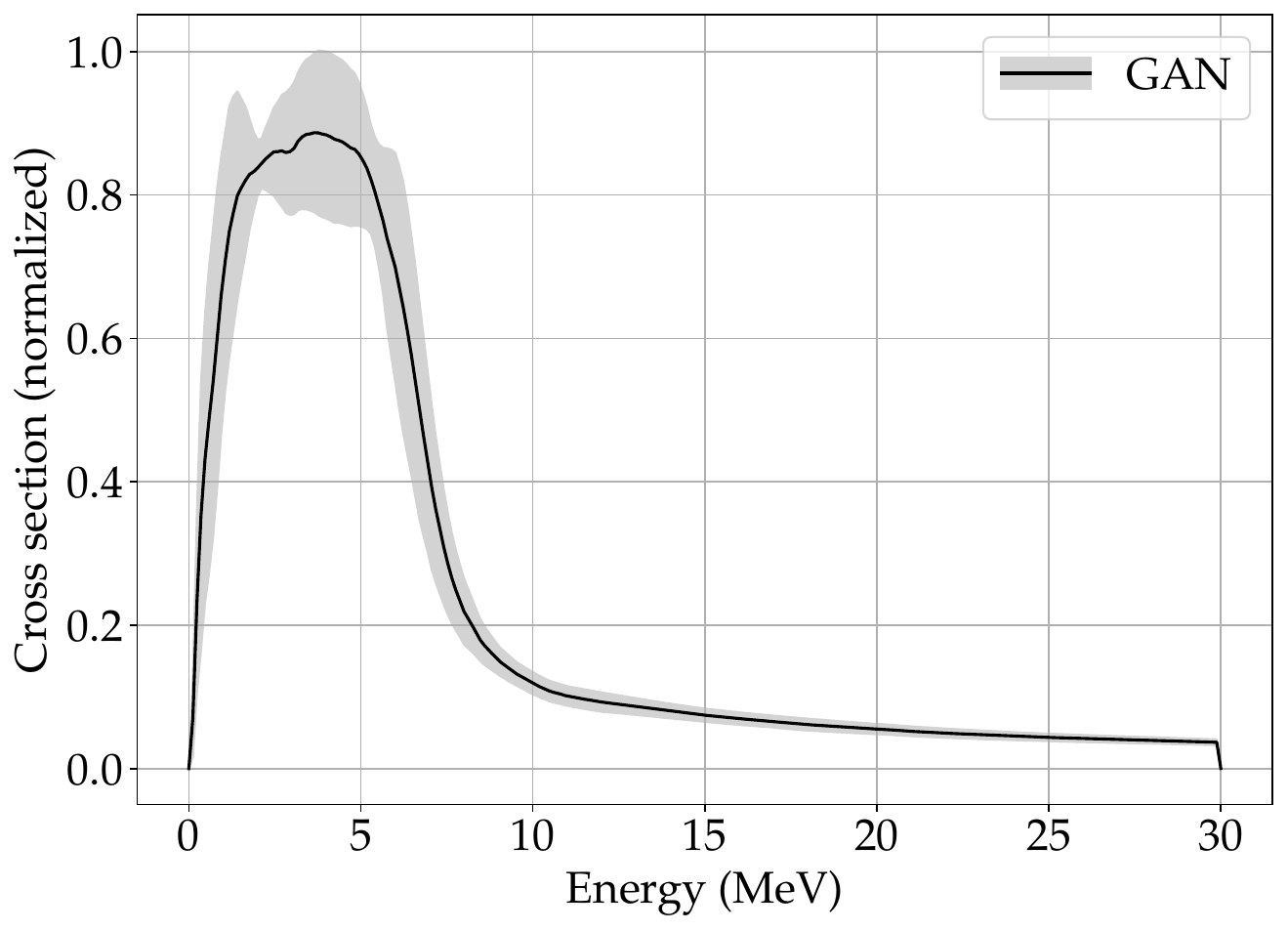}
\caption{Ensemble prediction of $^{158}$Ce represented by a median and standard deviation band, which is computed using a weighted average of curves in the ensemble.}
\end{subfigure}
\caption{Ensemble prediction of the (normalized) cross section of Cerium-158, using all possible paths within a bounded local region of the chart of nuclides. TENDL does not include an evaluation for this cross section, so this prediction may be considered an extrapolation. }
\label{fig:ml:ensemble_prediction_158Ce}
\end{figure*}

\begin{figure*}
\begin{subfigure}{\textwidth}
\includegraphics[scale=0.4]{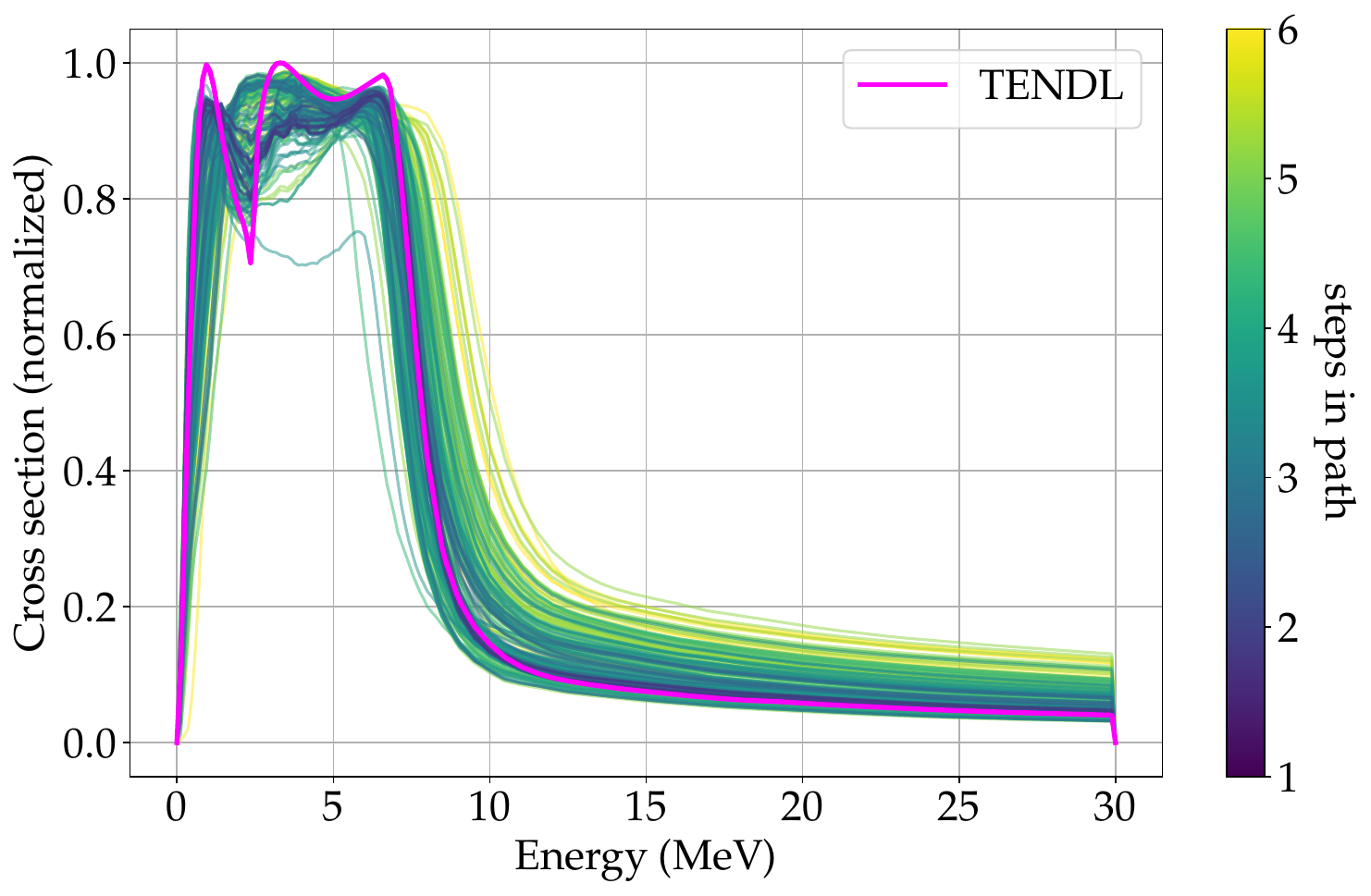}
\caption{Ensemble prediction of $^{158}$Sm with all 213 predicted curves shown. GAN predictions are shown in viridis (blue-green-yellow) where the color corresponds to the number of linked translations.  }
\end{subfigure}
\bigskip
\begin{subfigure}{\textwidth}
\includegraphics[scale=0.4]{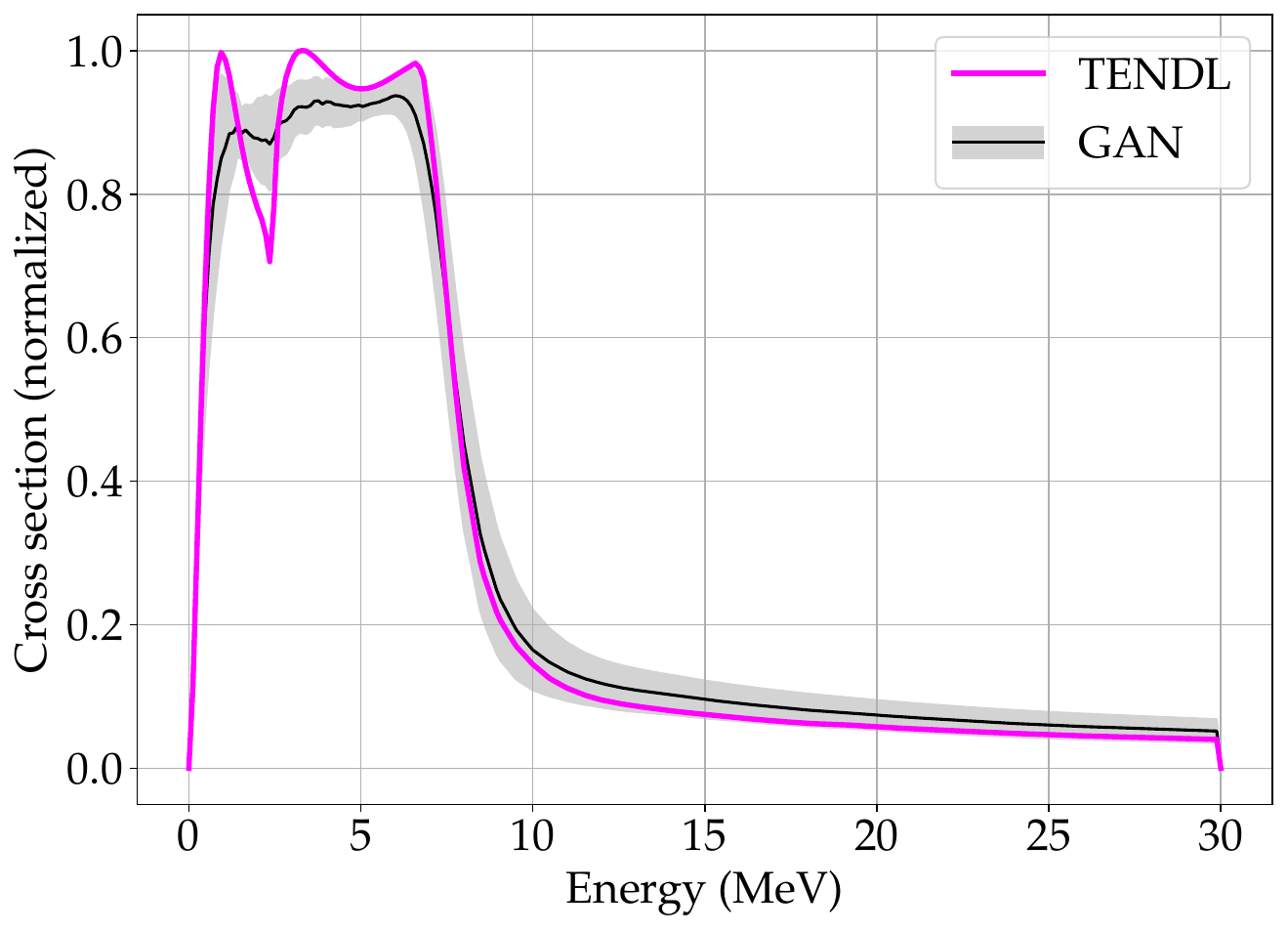}
\caption{Ensemble prediction of $^{158}$Sm represented by a median and standard deviation band, which is computed using a weighted average of curves in the ensemble.}
\end{subfigure}
\caption{Ensemble prediction of the (normalized) cross section of Samarium-158, using all possible paths within a bounded local region of the chart of nuclides. While the ensemble reproduces the curve with small errors, the strong resonance around 2 MeV is only predicted by fewer than half of the ensemble curves.}
\label{fig:ml:ensemble_prediction_158Sm}
\end{figure*}

\begin{figure*}
\begin{subfigure}{\textwidth}
\includegraphics[scale=0.4]{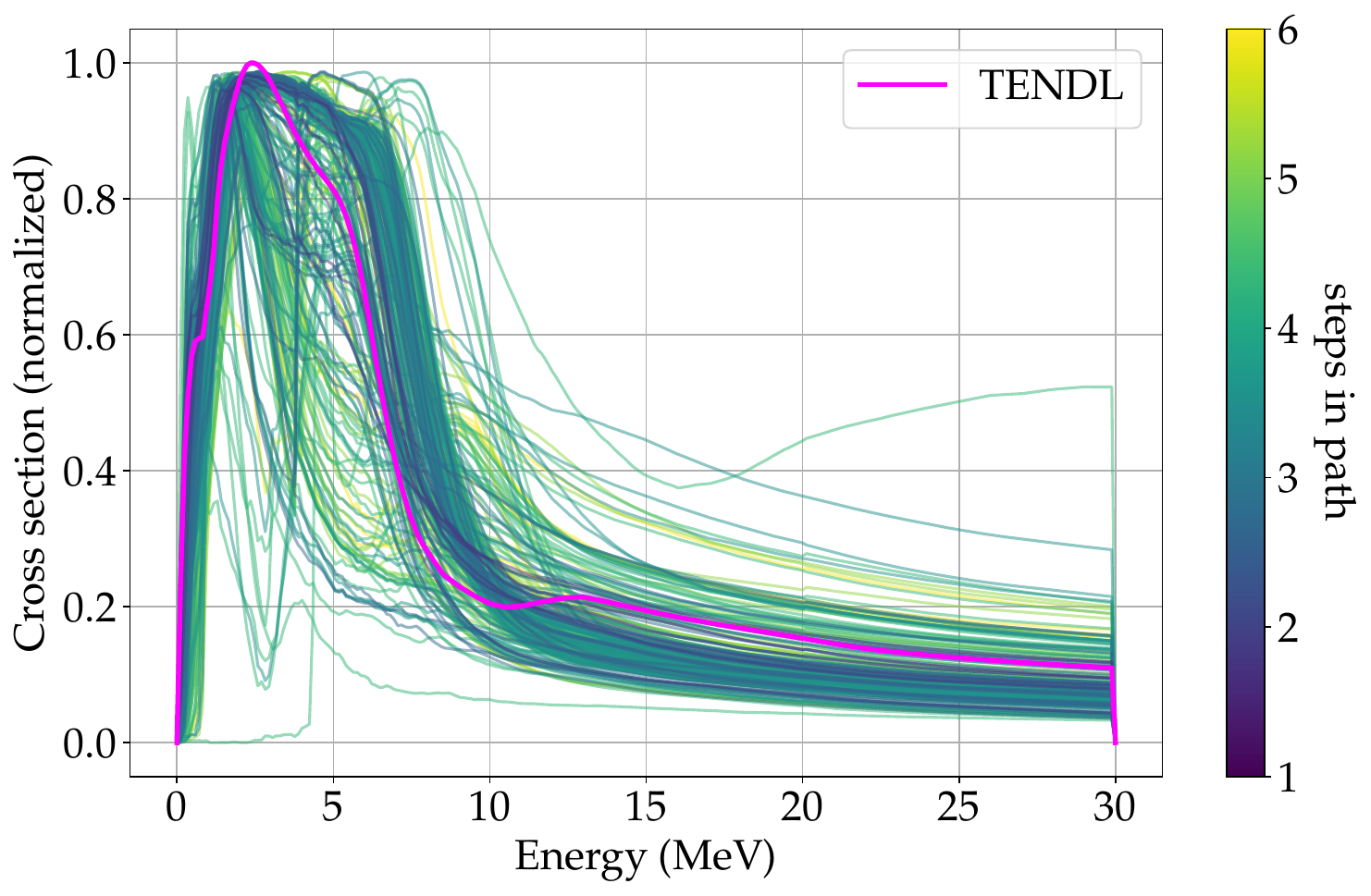}
\caption{Ensemble prediction of $^{234}$U with all 272 predicted curves shown. GAN predictions are shown in viridis (blue-green-yellow) where the color corresponds to the number of transform applications (i.e. length of path across chart). }
\end{subfigure}
\bigskip
\begin{subfigure}{\textwidth}
\includegraphics[scale=0.4]{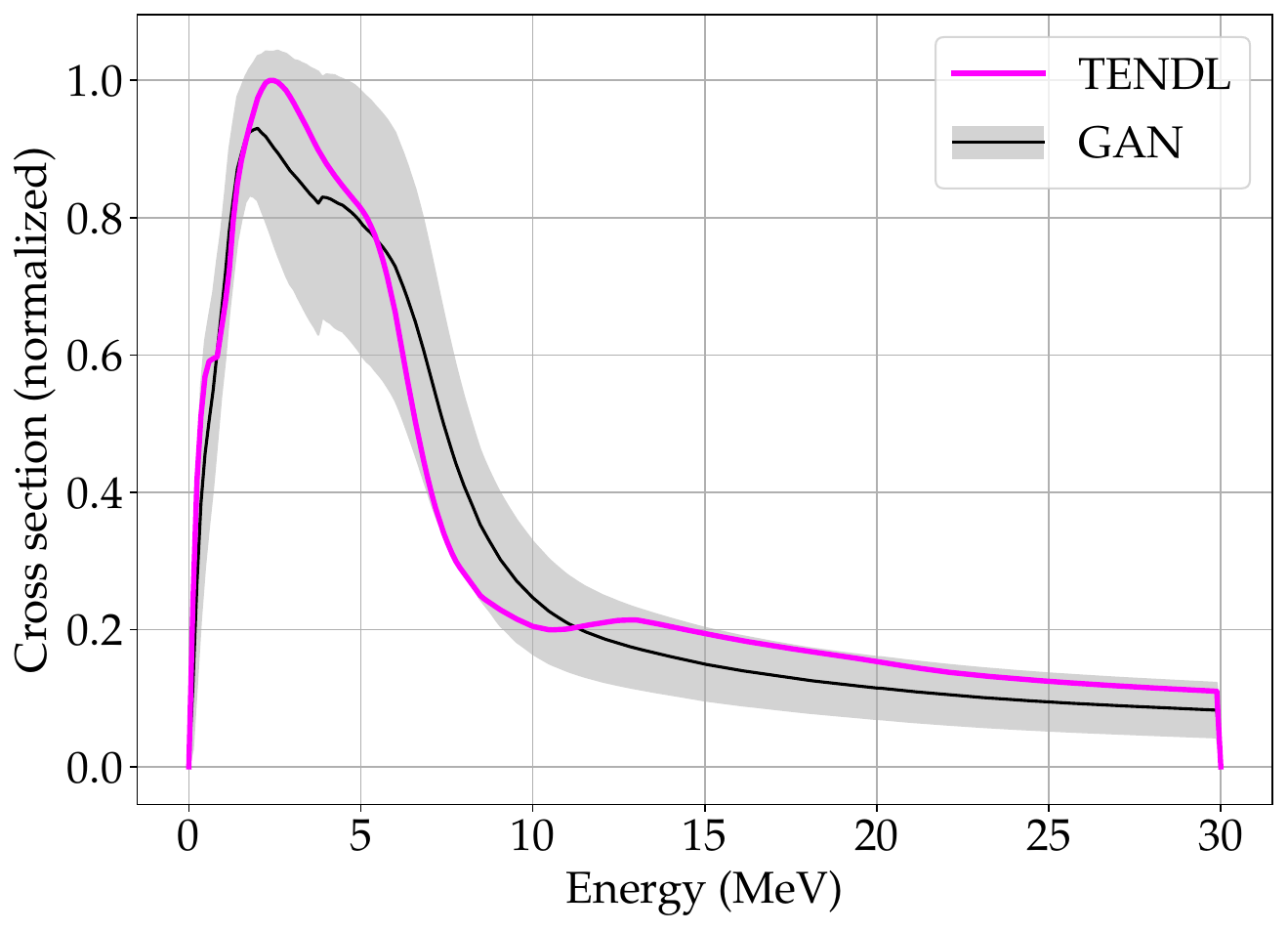}
\caption{Ensemble prediction of $^{234}$U represented by a median and standard deviation band, which is computed using a weighted average of curves in the ensemble. }
\end{subfigure}
\caption{Ensemble prediction of the (normalized) cross section of Uranium-234, using all possible paths within a bounded local region of the chart of nuclides. 
Larger errors are expected for very heavy (and very light) nuclides, and significant variability can be seen in the individual predictions. Averaging, however, shows the true cross section indeed lies within the confidence band.}
\label{fig:ml:ensemble_prediction_234U}
\end{figure*}

Examples of ensemble predictions are shown in Figures \ref{fig:ml:ensemble_prediction_88Sr}, \ref{fig:ml:ensemble_prediction_94Zr}, \ref{fig:ml:ensemble_prediction_158Ce},  \ref{fig:ml:ensemble_prediction_158Sm}, and \ref{fig:ml:ensemble_prediction_234U}. Each figure shows the TENDL cross section evaluation in pink and the predicted ensemble in thin, colored lines. In the upper (a) plots, the line color corresponds to the number of steps in the path leading to that prediction (see color bar). In lower (b) plots, we show the results of a Gaussian weighted average with predictions weighted by the inverse square of the path length (in the number of steps), so predictions with more linked model evaluations are discounted. This model-averaging technique is by no means rigorous and is included here primarily to aid visual interpretation.
How exactly the ensemble of curves should be transformed into a single estimate of the cross section is nontrivial, and the solution (possibly Bayesian model-mixing \cite{hoeting_bayesian_model_averaging}) is beyond the scope of this paper. 
Figures \ref{fig:ml:ensemble_prediction_88Sr} and \ref{fig:ml:ensemble_prediction_94Zr} show good predictions and a reasonable spread of errors. As can be seen in Fig.~\ref{fig:ml:ensemble_prediction_94Zr}, short paths are generally more accurate than long paths. This makes sense since predictions from longer paths might accumulate errors that compound with each application. 

Fig.~\ref{fig:ml:ensemble_prediction_158Ce} shows predictions for Cerium-158, which does not have a cross section evaluation in the TENDL library. 
Interestingly, we see the ensemble prediction is bimodal; that is, each prediction is centered around one of two curves. 
The first mode increases very fast at 0 MeV, then has a shorter secondary peak around 6 MeV (which is a very common feature in the inelastic neutron scattering channel). 
It corresponds to predictions from shorter paths (blue and green curves), so it is likely more accurate. 
The second mode, which only has one major peak around 5 MeV, is created by longer paths (yellow and red curves) and thus is likely not as accurate as the other. 
We can thus inspect the ensemble result and glean a prediction for the cross section of Cerium-158, which does not have experimental measurements. 

Lastly, Fig.~\ref{fig:ml:ensemble_prediction_234U} illustrates an ensemble prediction with large variability. Ensemble predictions for heavy nuclides, like Uranium-234, are likely not as accurate as those for lighter nuclides, partly because there are fewer data points for heavy nuclides in the training data (which can be seen as the chart gets thinner at the heavy end, nuclides have fewer around them). 
Furthermore, the $^{238}$U cross section evaluation has been carefully tuned and thus may break from local systematic trends learned by the GAN.
A future priority for this research may be to decide how to better summarize these ensemble predictions with large variance.

\section{\label{sec:conclusion}Conclusion}

We have developed a predictive AI system to study intricate systematic trends in nuclear cross section evaluations in a subset of the TENDL library. 
This is, to our knowledge, the first instance of deep generative learning being used to predict cross section evaluations.

Using well-established machine learning models, namely the variational autoencoder (VAE) and cycle-consistent generative adversarial network (cycleGAN), we have constructed a novel ensemble system capable of leveraging changes in cross section as a function of proton and neutron numbers of the target nuclide.
The VAE provides a reliable compression scheme, reducing the size of cross section data by a factor of 8 and mapping to a smooth distribution.
The cycleGAN used in this project is originally based on other models that have shown substantial success in image-to-image translation, however, the cycleGAN used here is truly a novel development, tailored specifically to its task.
The more powerful networks of the cycleGAN are responsible for learning intricate changes in the cross section itself, but it also uses a smaller discriminator network to ensure the correlations between cross sections match those present in training data.
Furthermore, the ensemble nature of the system means that it is very easy to assign uncertainty bands to predictions.

Although the system has some weaknesses, such as sensitivity to outliers in training data, it also exhibits remarkable robustness and stability in predictions of many cross sections at once. 
The linked translations shown in Figs.~\ref{fig:ray_extrap_1},\ref{fig:ray_extrap_2},\ref{fig:ray_extrap_3} illustrate the remarkable ability of this model to allow for a range of discrepancy in inputs while still producing sensible results. 
Leveraging this stability in further development could allow for valuable predictions to be made of cross section curves beyond the limits of current experimental techniques. 
Although the present model may not be appropriate for larger problems, such as targeting correlations between reaction channels, we hope this research can inspire other researchers to bring creativity and insight to this problem.

\section{Acknowledgments}
This work was performed under the auspices of the U.S. Department of Energy by Lawrence Livermore National Laboratory under Contract DE-AC52-07NA27344, supported in part through the Laboratory Directed Research and Development under project 23-SI-004. JF is supported by the U.S. Department of Energy, Office of Science, Office of Nuclear Physics contract DE-AC02-06CH11357, as well as the SCGSR program, the NUCLEI SciDAC program, and Argonne LDRD awards.  Computing support came in part from the LLNL institutional Computing Grand Challenge program.

\appendix

\section{\label{app:simplegans}A little background on generative adversarial networks}

Generative models are a family of machine learning techniques designed to learn a probability distribution.
The variational auto-encoder discussed in this paper is a type of generative model: the decoder network can be fed a normal random variable and generate random samples of the target distribution. 
What sets the GAN apart is a modification of the generator network's loss function: rather than relying on the latent space distribution of data, one introduces an entirely separate network to identify definitive trends in training data and penalizes the generating network when the output does not match those trends. 
As such, the adversarial nature of the model allows for a generalization of the statistical requirements present for the VAE.

The simplest GANs consist of two networks: one generator $G$ and one discriminator $D$, both acting on data $x \in X$.  
The generator learns a mapping to the target distribution, and the discriminator scores the validity of generated data, $D(x') = p(x' \in X | x \in X)$.
This type of GAN is useful for fully unsupervised learning; that is, the generator learns the relevant probability distribution without any labeled inputs (aside from the labels for $x \in X$ and $G(x) \not\in X$, which are supplied to the discriminator, but these are trivial to generate). 

During training, the weights of networks $G$ and $D$ are optimized with respect to a loss function, and the networks compete with one another in what is notoriously tantamount to finding a Nash equilibrium. Methods for reliable GAN training is an open area of research. 
Training the system results in a generator that can ``convincingly'' (according to the discriminator) create new samples $x'$, the correlations of which closely resemble those of the training data. 
We can introduce the notation $x' \simin X$, meaning $x'$ closely resembles an element of the set $X$, or similarly $x'$ looks like a sample of the distribution from which all $x \in X$ could be drawn. 
That is, formally, $x' \not{\sim} P(X)$, but $p(x' \sim P(X))$ is nowhere near zero. 
We may denote this simple GAN as the pair $\{G, D\}$ and write the networks as explicit functions of weights $\theta$ as $G=G(\theta_G)$ and $D=D(\theta_D)$ for clarity.

The loss function for the simple GAN is different for the generator and discriminator, but both can be expressed in terms of the following two expectation values ($\mathbb{E}$). 

\begin{equation}\label{eq:gan_loss_formal}
    v =   \mathbb{E}_{x \sim P(X)} \log( D(x) )  +  \mathbb{E}_{x \sim P(G)} \log ( 1 - D(x))
\end{equation}

The distribution $P(G)$ means the distribution of the generator samples. The respective loss functions are $\mathcal{L}_G (\theta_G) = v(\theta_G)$ and $\mathcal{L}_D (\theta_D) = - v(\theta_D)$. The discriminator loss is low when it correctly identifies both true data and generated data, thus both terms are large. The generator loss only depends on the second term and is low when the discriminator assigns a high probability to the generator outputs; thus, we might say the discriminator is being ``deceived'', classifying the generated data as real. 

Although finding a true Nash equilibrium is ideal, this can be very difficult in practice. 
If both networks are training, that which has the advantage will fluctuate, but this process can continue for a long time without a clear indication of making progress.  
An approximation may be found by first training a discriminator to effectively classify the data, then subsequently training the generator by itself. 
So, it is common for GANs to be trained this way, repeating the process to produce a better approximation to equilibrium.

In practice, we may express this loss function in Eq.~\ref{eq:gan_loss_formal} in a slightly different way. One often uses a loss function like binary cross-entropy (BCE) to evaluate probabilities, which the discriminator network emits. BCE is the preferred loss function when the classifier network ends with a single sigmoid neuron: it compares a label probability $y$ with a predicted probability $p$ and produces one number that decreases as $y \rightarrow p$.  (In practice, one may design a classifier that emits a so-called \textit{logit} value on $(- \infty, + \infty )$ and a modified version of the BCE to handle logits instead of probabilities. This is mathematically equivalent but might allow for better convergence in some cases.)

\begin{equation}\label{eqn:bce}
    BCE\left[p,y\right] := - \left[ y \log(p) + (1-y) \log(1-p) \right]
\end{equation}

BCE can also be evaluated for a set of inputs, as in batch training, simply by averaging the individual BCE values for each pair.  In terms of BCE for a single batch of $N_b$ data points, the adversarial loss function is the sum of $\mathcal{L}_D$ and $\mathcal{L}_G$ in Eq.~\ref{eq:gan_losses}.

\begin{equation}\label{eq:gan_losses}
    \begin{aligned}
        \mathcal{L}_D (\theta_D) &= \frac{1}{N_b} \sum_{i=1}^{N_b} \left[  BCE[D(x_i),1] + BCE[D(G(z_i)),0] \right] \\
        \mathcal{L}_G (\theta_G) &= \frac{1}{N_b} \sum_{i=1}^{N_b} BCE[ D(G(z_i)),1]
    \end{aligned}
\end{equation}

Ultimately, the GAN learns to approximate the probability distribution of the data $P(X)$, and the evaluation of the generator on random noise $G(z)$ approximates sampling from $P(X)$. 
To better adapt the model to our nuclear data problem, we consider a slightly more advanced form called the \textit{cycle-consistent} GAN.  

Zhu et al. \cite{cyclegan_zhu} demonstrated the effectiveness of cycle-consistent GANs (cycleGANs) for image-to-image translation: two GANs $\{G, D_G\}$ and $\{F, D_F\}$, may be used to map between two distinct probability distributions, for $x \in X$ and $y \in Y$, as $G(x) = y' \simin Y$ and $F(y) = x' \simin X$ (read: $y'$ is similar to elements of $Y$, etc.). The discriminator networks $D_G$ and $D_F$ score the validity of the two resulting candidates: $D_G(y') = p(y' \in Y | y \in Y)$ and $D_F(x') = p(x' \in X | x \in X)$. 
The cycleGAN loss functions add two new terms to the loss functions in Eq.~\ref{eq:gan_losses}. First, cycle loss ensures that applying the generators in succession returns the input (i.e., the convolution of co-inverse generators is the identity function). For a batch of $N_b$ data points, cycle loss is as follows.

\begin{equation}\label{eqn:cycle_loss_appendix}
    \mathcal{L}^G_\text{cycle} = \frac{1}{N_b}\sum_i^{N_b} |X_i - F(G(X))_i| + \frac{1}{N_b}\sum_i^{N_b} |Y_i - G(F(Y))_i| 
\end{equation}

One such application is \textit{style transfer} \cite{stylegan_nvidia} where $G$ learns to transform $X$ so that correlations match those of $Y$, and $F$ learns the reverse.  Style transfer employs the constraint that generators be idempotent ($G^2=G$): $G(y) \simin Y $ and $F(x) \simin X$ because the generators should act as projections into the respective spaces (that is, $G(x) \simin Y$ but $G(G(x)) \simin Y$ as well). The term ``cycle-consistent'' refers to an additional constraint that each generator be the inverse of its partner: $G(F(y)) = G(x') = y'$ and $F(G(x)) = F(y') = x'$.  
This is one of several mechanisms that encourage the generators to learn systematic trends rather than simply ``memorize'' the training data (e.g., network configurations that only serve one purpose). 

\section{\label{app:challenges}Challenges with the convolutional GAN}

Our initial attempts at this project involved not the separate convolutional VAE and dense GAN but rather a single convolutional GAN, which was intended to learn both latent space encodings and translation of cross sections. 
This construction was difficult to work with, and although it is not clear why, we may learn some things by comparing that model to this final version. 
In the old model, the generator of the cycleGAN followed a U-net design \cite{unet_ronneberger}, which is very successful when applied to image datasets. 
The U-net design has shown success in GAN models on 2D image data, so our assumption was that 1D cross section data is similar enough that we could use a 1D version of the same model structure. 
The loss function was the same as that of our dense GAN; the discriminator network followed the design of a simple multi-layer convolutional classifier. However, when training it, we found the U-net model to have poor convergence properties. 
It was especially difficult to control overfitting while maintaining good predictions and optimizing the coefficient of target loss. 
If the coefficient $\lambda_\text{pred}$ (see Eq.~\ref{eqn:total_loss}) is too large, then the model would overfit, and predictions would be unreliable; if it were too small, the model would not converge (i.e., predicted cross sections were very noisy and would not match the training set).  

\begin{figure*}
\includegraphics[scale=0.45]{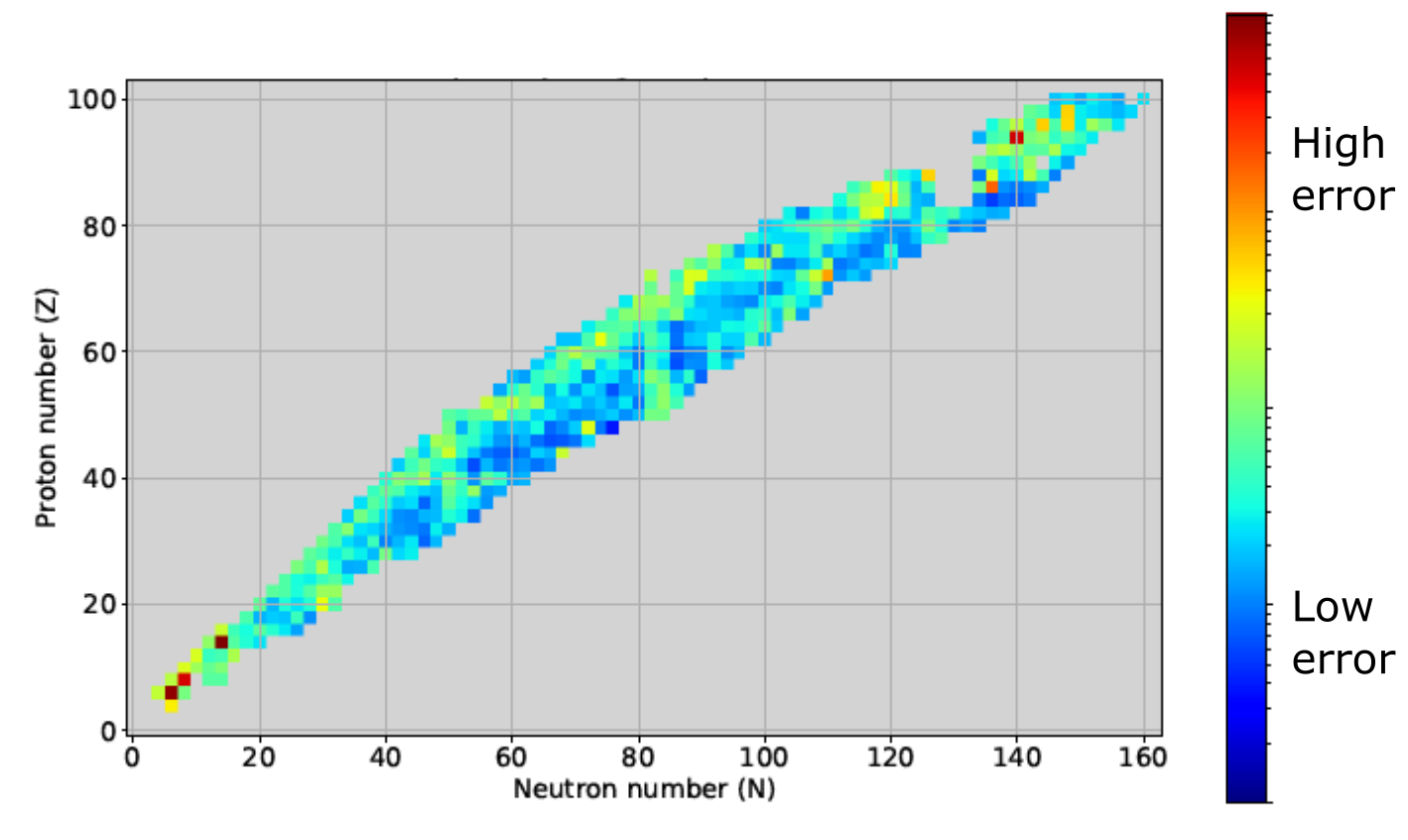}
\caption{Average error of local predictions using the U-net GAN model, which was prone to overfitting, unstable, and ultimately abandoned in favor of the VAE+GAN system. Note that higher errors are clearly visible around magic numbers $N=50,82$, indicating the model has not learned how cross sections are dependent on shell structure. The nuclides with very high error (red) are those where the cross-section is far more detailed than its neighbors (e.g., O$^{16}$). }\label{fig:chart_heatmap_old}
\end{figure*}

A reassuring physics connection can be seen in comparing the results of the U-net with the one presented. Since the U-net would not converge fully, it could not learn systematic trends across the chart, and in particular, Fig.~\ref{fig:chart_heatmap_old} shows that error increases around $N=50$ and 82, which are magic numbers. 
This indicates that the model had not learned to incorporate changes in the cross sections that are due to shell structure; such errors are not present in the final version of our model.

All this is not to say that a 1D U-net style cycleGAN is a poor design in general, only that in our particular implementation for this problem, it did not give good results. Whether that model could be successful in some nuclear data applications may be a worthy question to pursue, but that is beyond the scope of this research.

\bibliography{master}

\end{document}